\documentclass{emulateapj}

\usepackage{graphicx}
\usepackage{apjfonts}
\usepackage{longtable}
\usepackage{natbib}

\newcommand{\Msun}[0]{$M_{\sun}$}

\begin{document}

\title{Lithium Depletion of Nearby Young Stellar Associations}

\author{Erin Mentuch, Alexis Brandeker\altaffilmark{1}, Marten H. van Kerkwijk, Ray Jayawardhana}
\affil{Department of Astronomy \& Astrophysics, University of Toronto, 50 St.\,George Street, Toronto, Ontario M5S\,3H4, Canada}
\and
\author{Peter H. Hauschildt}
\affil{Hamburger Sternwarte, Gojenbergsweg\,112, 21029, Hamburg, Germany}

\altaffiltext{1}{Current address: Stockholm Observatory, AlbaNova University centre, SE-106\,91 Stockholm, Sweden}

\begin{abstract}

We estimate cluster ages from lithium depletion in five
pre-main-sequence groups found within 100\,pc of the Sun: TW~Hydrae
Association, $\eta$~Chamaeleontis Cluster, $\beta$~Pictoris Moving
Group, Tucanae-Horologium Association and AB~Doradus Moving Group.  We
determine surface gravities, effective temperatures and lithium
abundances for over 900 spectra through least squares fitting to
model-atmosphere spectra.  For each group, we compare the dependence
of lithium abundance on temperature with isochrones from
pre-main-sequence evolutionary tracks to obtain model dependent ages.
We find that the $\eta$\,Chamaelontis Cluster and the TW~Hydrae
Association are the youngest, with ages of $12{\pm6}$\,Myr and $12{\pm8}$\,Myr,
respectively, followed by the $\beta$~Pictoris Moving Group at $21{\pm9}$\,Myr, the
Tucanae-Horologium Association at $27{\pm11}$\,Myr, and the AB Doradus
Moving Group at an age of at least $45$\,Myr (where we can only set a
lower limit since the models -- unlike real stars -- do not show much
lithium depletion beyond this age).  Here, the ordering is robust, but
the precise ages depend on our choice of both atmospheric and
evolutionary models.  As a result, while our ages are consistent with
estimates based on Hertzsprung-Russell isochrone fitting and dynamical
expansion, they are not yet more precise.  Our observations do show
that with improved models, much stronger constraints should be
feasible: the intrinsic uncertainties, as measured from the scatter
between measurements from different spectra of the same star, are very
low: around 10\,K in effective temperature, 0.05\,dex in surface
gravity, and 0.03\,dex in lithium abundance.

\end{abstract}

\section{Introduction}\label{s:intro}

Triggered largely by the discovery of young stars in the {\em ROSAT} X-ray Satellite
All-Sky Survey, over the last decade several nearby pre-main-sequence
(PMS) star groups have been identified (for a review, see 
\citealt{zuc04}).  Ranging in age from roughly 6\,Myr to $\sim$100\,Myr, five
separate groups can be distinguished: TW Hydrae Association (TWA),
$\eta$~Chamaeleontis cluster ($\eta$\,Cha), $\beta$~Pictoris Moving
Group (BPMG), Tucanae-Horologium association (TUCHOR) and AB~Doradus
Moving Group (ABD).  The common space motions and localized sky positions 
suggest that these groups are likely connected to the Sco-Cen star
forming region located $\sim$100\,pc away in the southern
hemisphere.  \citet{mam99} and \citet{son03} have traced back the space
motion of members in BPMG, TWA and $\eta$\,Cha and argue that the
groups are related to a star formation burst in the
Sco-Cen region as a result of the passing of the Carina arm
$\sim$60\,Myr ago. 

These groups, because of their close vicinity, are excellent
laboratories for studying star and planet formation.  Well constrained
ages are necessary to make conclusions about timescales of, e.g., disk
dissipation and planet formation.  Already, observations 
from the same sample presented
in this paper have revealed that accretion disks can last up to
$\sim$10\,Myr, but beyond this is rare \citep{jay06}.

\defcitealias{bar98}{BCAH98} 
Previously, ages have been derived from Hertzsprung-Russell (HR)
diagram fitting, group dynamics and lithium abundance
measurements. \citet{luh04} provide an HR diagram isochrone age for
$\eta$\,Cha of $6^{+2}_{-1}$\,Myr derived from the evolutionary
models of \citeauthor{bar98} (\citeyear{bar98}; hereafter \citetalias{bar98})  and \citet{pal99}, which agrees well with the
dynamical expansion age of 6.7\,Myr determined by \citet{jil05}. The
dynamical age of TWA has been harder to determine because of
inconsistent space motions among its more than 30 members. An inferred
age of $8.3\pm0.8$\,Myr is given to TWA based on the dynamical motion
of four members \citep{del06}. However, the likely complex dynamical
evolution of TWA has led to several plausible evolutionary
scenarios. \citet{mak05} attribute this complex evolution to a chance 
encounter with Vega, while \citet{law05} suggest
TWA is composed of two separate groups, based on bimodal rotation
period distributions with distinctly separate ages of $\sim$\,10\,Myr
and $\sim$\,17\,Myr. More recently, \citet{bar06} finds a conservative
age of $10^{+10}_{-7}$\,Myr by comparing ages from HR diagram
isochrone comparisons (from BCAH98) and lithium abundances.

The slightly older group BPMG has an estimated age of $12^{+8}_{-4}$\,Myr 
based on HR diagram isochrone comparisons (from \citetalias{bar98}) and lithium abundances 
\citep{zuc01}, with three dimensional motions that are consistent 
with a dynamical expansion age of 11.5\,Myr \citep{ort02}. \citet{fei06}
independently derived an age of $13^{+4}_{-3}$\,Myr for the recently 
confirmed wide binary system of 51~Eri and GJ\,3305, part of BPMG.

Known to be older than BPMG, but younger than the Pleiades, TUCHOR
has an age of 20--40\,Myr based on H$\alpha$ measurements, X-ray
luminosity, rotation and lithium abundances in comparison to other
young clusters like TWA and the Pleiades \citep{zuc00,ste00}.
 
Perhaps, the most debated age is that of the ABD group. \citet{zuc04b}
derived an age of 50\,Myr by comparing the H$\alpha$ emission strength of
ABD members to members of the younger TUCHOR association, in addition
to fitting its three M-type members to HR diagram isochrones. In
contrast, \citet{luh05} compared HR isochrones of ABD members to those 
of two well-observed clusters with ages of 50 and 125\,Myr and suggested
that the ABD group is coeval with the Pleiades at an age of 100--125\,Myr.
 The latter age is strongly supported by \citet{ort07},
who compute full 3D galactic orbits of ABD and the Pleiades
cluster and show the dynamics of the two groups can be traced back to
a common origin of $119\pm20$\,Myr ago.

A relatively new approach to age estimates is to use the evolution of
the lithium abundance for low-mass, partially and fully convective PMS
stars \citep{bil97,jef05}. The initiation and duration of
lithium depletion in PMS stars is dependent on mass and is very
sensitive to the central temperature. Lithium is converted into helium
in $p,\alpha$ reactions in cores of low-mass stars when the
temperature reaches $2.5\times10^{6}$\,K. The lower the stellar mass,
the longer the time it takes to reach this critical temperature. For
example, a 0.6\,\Msun\ star begins to burn lithium at an age of
3\,Myr, while a lower mass star at 0.1\,\Msun\ begins to burn lithium
at an age of 40\,Myr. Stars with $M<0.06$\,\Msun\ never reach this
typical temperature, while stars with 0.6\,\Msun\ $<M<$
1.2\,\Msun\ burn lithium for a short period (1--2\,Myr) until a
radiative core develops, and more massive pre-main sequence stars do
not destroy lithium in their envelope at all. The result of these
processes is a dip in the lithium abundance as a function of luminosity
(and consequently, a function of effective temperature),
only affecting stars with spectral types late than F5. As a group
ages, this dip becomes deeper and widens on the cool end as
progressively cooler stars reach the critical core temperature.

The cool end of this dip has been used to date coeval groups
containing late M-dwarf stars by identifying the lithium depletion
boundary (LDB). The LDB marks the
luminosity above which all stars will
have depleted their lithium. The lithium is very quickly depleted in
these low-mass stars, so the LDB marks a sharp jump from initial to
near depleted lithium abundances. As the temperature in the cores of
PMS stars increases in time, the LDB will shift to cooler temperatures
as a cluster ages. LDB ages have been determined for the Pleiades (125
$\pm$ 8 Myr), $\alpha$\,Per (90 $\pm$ 10 Myr), IC\,2391 (53 $\pm$ 5
Myr), and NGC\,2547 (35 $\pm$ 4 Myr)
\citep{sta98,bar99,sta99,bar04,jef05}.

In this paper, we fit more than 900 multi-epoch, high-resolution spectra, introduced
in \S\ref{s:obs}, of 121 low-mass PMS stars to synthetic spectra
created from PHOENIX model atmospheres (see \S\ref{s:models}). We
begin by identifying the model with the best-fitting surface gravity, $\log
g$, and effective temperature, $T_\mathrm{eff}$ for each 
observed spectrum, as described in
\S\ref{s:gravtemp}. In \S\ref{s:Li}, we use these
best-fit $T_\mathrm{eff}$ and $\log g$ to find the lithium abundance
by fitting model spectra to both the 6\,104\,\AA\ and
6\,708\,\AA\ lithium features in the observations, but now using
lithium abundance as a free parameter.  For comparison, we also measure the
equivalent width (EW) of the 6\,708\,\AA\ lithium doublet which allows us
to chronologically order the groups based solely on empirical 
measurements. We proceed by comparing the
measured lithium abundance distribution to that predicted by PMS
evolutionary models of \citetalias{bar98} and \citet{sie00} and
estimate model dependent ages for each of the PMS groups.

\section{Observations}\label{s:obs}

High-resolution spectra were obtained during six separate observation
runs, on a total of 19 nights, between December 2004 and April 2006,
utilizing the MIKE spectrograph at the Magellan-Clay 6.5\,m telescope
on Las Campanas, Chile. MIKE is a double echelle instrument, covering
two separate wavelength regions. For this study, the red region from
4\,900\,\AA\ to 9\,300\,\AA\ is used. The raw data were bias subtracted 
and flat-fielded, and before extraction, the
scattered background in the spectrograph was subtracted by fitting splines
to interorder pixels. The spatial direction of the projected slit produced
by MIKE is wavelength dependent and not aligned with the CCD. We therefore
extracted spectra using a custom procedure develped in ESO-MIDAS, that
takes into account the tilt and optimally extracts the spectrum by
iteratively estimating the slit illumination function. For wavelength
calibration, exposures of a Thorium-Argon lamp were used, as well as
observed telluric absorption lines. A more detailed account of the
reductions and a log of the observations will appear in a forthcoming paper
(A.\ Brandeker et al.\ 2008, in prep.). Multiple spectra for many 
of the targets in our sample were taken in
order to search for multiplicity and perform variability studies.

With no binning and using the 0\farcs35 slit, the spectra in this
study have a resolution of R$\sim$60\,000. The pixel scale was
0\farcs13\,pix$^{-1}$ in the spatial direction, and about
24\,m\,\AA\,pix$^{-1}$ at 6\,500\,\AA\ in the spectral
direction. Integration times were chosen so that the signal-to-noise
ratio (S/N) $\gtrsim$70 per spectral resolution element at
6\,500\,\AA, except for the brightest stars where this would have
implied an exposure shorter than 120\,s. In those cases, we used the
longest exposure time shorter than 120\,s that did not saturate the
detector, giving (S/N) = 70--500, depending on seeing.

For this study, we only use objects with spectral types later than F5,
and earlier than M5.  Hotter stars do not show depleted lithium
abundances in their atmospheres, because their cores are already
radiative when lithium burning starts. Cooler stars were not included
in the survey because they were too faint.  We also exclude obvious
spectroscopic binaries. In total we have a sample of 121 stars, with
11 from $\eta$\,Cha, 32 from TWA, 23 objects from BPMG, 35 
from TUCHOR, and 22 from ABD. For comparison, we also analysed 
20 radial-velocity standard stars from the field.

\section{Models}\label{s:models}

We compute synthetic spectra using version 14 of the PHOENIX model
atmosphere package \citep{hau98}. The model atmospheres are described
in \cite{kuc05,kuc06}. In total, we have 240 model spectra
covering a spectral region from 3\,200\,\AA\ to 10\,000\,\AA\, with
0.03\,\AA\ spectral resolution, with temperatures ($T_\mathrm{eff}$)
ranging from 2\,500\,K to 6\,500\,K in steps of 100\,K, and surface
gravities ($\log g$) ranging from 3.0 to 6.0 in steps of 0.5
(i.e.\ gravities ranging from $10^3$\,cm\,s$^{-2}$ to
$10^6$\,cm\,s$^{-2}$).  For each of these temperature and gravity
combinations, we have additional models with different lithium
abundances ($N_{\rm Li}$) ranging from $\log N_{\mathrm{Li}}=0.0$ to
4.0 in steps of 0.5 (where the normalization is such that $\log
N_{\mathrm{H}}=12$), which cover the 6\,104\,\AA\ and
6\,708\,\AA\ lithium absorption lines. All models were calculated at solar
metallicity.

The atmospheric models are calculated under the assumptions 
 of local thermodynamic equilibrium (LTE). \citet{car94} have systematically
 analyzed the effects of non-LTE on the formation of the \ion{Li}{1} line in 
 cool stars. They show that that the non-LTE effects can lead to discrepancies
 in lithium abundances measured from the 6\,708\,\AA\ lithium doublet
  from about -0.3 to 0.3 dex depending on the depletion
 and temperature. We attempt to minimize the exclusion of non-LTE 
 effects by also fitting the 6\,104\,\AA\ lithium feature whose non-LTE
 corrections are opposite to those of the 6\,708\,\AA\ line \citep{car94}.
 
 In addition, we do not consider the effects of 
chromospheric and/or magnetic activity. To compensate for the former, we exclude 
from our fitting method emission lines due to chromospheric activity
that are frequently seen in classical T~Tauri and post T~Tauri stars 
(described further in \S\ref{s:mods}). We also note here that  
very strong lines --- which are completely dominated 
by pressure broadening, a relatively poorly understood process ---
require special treatment to reproduce the line profile. Although the
 models include attempts to do this, it does not always work satisfactory.
The resulting mismatches primarily affects lines 
from neutral metals that are highly abundant and very optically 
thick (such as \ion{K}{1} and \ion{Na}{1}). For less abundant neutral 
metals, such as \ion{Li}{1}, the effects should be minimal.

\section{Surface Gravities and Effective Temperatures}\label{s:gravtemp}

We derive surface gravities and
effective temperatures, following the methodology of \citet{moh04}, 
by fitting theoretical spectra to
our observed spectra in small, 80\,\AA\ wide, spectral regions that
contain absorption lines or molecular bands that are highly sensitive
to surface gravity and/or effective temperature.

\subsection{Modifications to the data}\label{s:mods}

To ensure our measurements would not be influenced by telluric
absorption lines, or by stellar emission related to activity, we made
some alterations to the observed spectra before fitting. For the
telluric lines, we first used multi-epoch observations of one of our
stars ($\beta$\,Pic) to identify the stronger atmospheric lines and determine their
depths relative to the continuum. Next, for each spectrum, we marked
a fixed spectral width of data points around the center of each
telluric line as contaminated; these regions are ignored in all fits
done below. We used widths of the marked regions that depended on the
strength of the telluric line (as measured for $\beta$\,Pic): a region
of 0.2\,\AA\ was removed for relatively weak lines, which had a depth
less than 80\% of the continuum. Regions of 0.6\,\AA\ were removed for
line depths between 80\% and 88\%, while for the strongest, deeply
saturated absorption lines, 1.0\,\AA\ wide regions were removed.

As our sample consists of young stellar groups, many of the stars show
signatures of activity, with relatively narrow chromospheric emission
lines found in many of our targets, and strong and wide emission lines
indicative of accretion in a few others \citep{jay06,sch07}.  The
chromospheric activity causes emission features at a number of
wavelengths, with the strengths and locations depending on spectral
types and age \citep{eis90}. To mitigate the effects of activity, we
exclude 2.4\,\AA\ regions around lines that frequently are seen in
emission from classical T~Tauri and post-T~Tauri stars \citep{sta05}:
\ion{He}{1} at 5\,875.6\,\AA, \ion{Na}{1} at 5\,890.0 and
5\,895.9\,\AA, \ion{O}{1} at 8\,446.5\,\AA, and \ion{Ca}{2} at
8\,498.0, 8\,542.1 and 8\,662.1\,\AA.

Some stars also show signs of accretion, which leads to strong, wide
emission features, potentially leading to less accurate fits.
Since only three out of eleven $\eta$\,Cha and two out of 32 TWA
members show signs of accretion \citep{jay06}, this will affect
relatively few stars in the sample and in practice we do not find any effect of 
accretion on the uncertainties involved in constraining $T_{\mathrm{eff}}$, 
$\log g$ and $\log N_{\mathrm{Li}}$, except for the EW of one star 
(see \S\ref{s:ew}). There are no known accretors in any of the older 
PMS groups investigated in this study.

\subsection{Selection of spectral regions}\label{s:spec_regions}

We locate spectral regions that are
strongly sensitive to surface gravity and temperature. Both
\citet{tor93} and \citet{kir91} have compiled lists of prominent
absorption features useful for spectral classification of low-mass
stars. For all, we checked whether our model spectra (see
\S\ref{s:models}) indeed showed strong temperature and/or gravity
sensitivity, and selected only those that did.

\begin{deluxetable}{cccccc}
\tablecaption{\small{Spectral Regions Fitted}}
\tablecolumns{12}
\tablewidth{0pc}
\tabletypesize{\small}
\tablehead{
	\colhead{Line ID} 				&
	\colhead{$\lambda$ (\AA)} 		&
	\colhead{$\Delta\lambda$ (\AA)} 	&
	\colhead{T (K)}					&
	\colhead{Sens.}					
}
\startdata 
\ion{Na}{1}\tablenotemark{a}		& 5\,893		& 	5\,850--5\,930 	&	2\,500--4\,500	&	T, $\log g$	\\ 
\ion{Fe}{1}\tablenotemark{a} 		& 5\,893, 5\,898 	&				&				&				\\
TiO\tablenotemark{a} 			& 5\,847--6\,058 	&				&				&	T 			\\
\tableline
VO\tablenotemark{b}	 		& 7\,851--7\,973 	&	7\,900--7\,980 	& 	4\,000--6\,500 	&	T 			\\ 
CN\tablenotemark{a,b} 			& 7\,916, 7\,941, 7963 & 			&				&				\\ 
\tableline
\ion{Na}{1}\tablenotemark{b} 		& 8\,183, 8\,195 	&	8\,150--8\,230 	&	2\,500--3\,000 	&	$\log g$ , T 	\\
\tableline
TiO\tablenotemark{a,b} 			& 8\,432, 8\,442, 8\,452 & 8\,400--8\,480 	&	2\,500--6\,500 	&	T 			\\ 
\ion{Fe}{1}\tablenotemark{b} 		& 8\,440, 8\,468 	&				&				&				\\ 
\tableline
\ion{Ca}{2}\tablenotemark{a,b}	 	 & 8\,498, 8\,542 & 	8485--8565 	&	2\,500--6\,500 	&	T 			\\ 
VO\tablenotemark{b} 	 		& 8\,521, 8\,538 	&				&				&				\\ 
\enddata	
\label{tab:spectralregions}
\tablecomments{A list of the spectral regions selected for our fitting 
method as described in \S\ref{s:spec_regions} \& \S\ref{s:methods}. 
The first and second columns identify and locate spectral features within the 
chosen spectral range that show strong sensitivity to temperature and/or 
surface gravity. The third column lists the 80\,\AA\ region that was selected in
our fitting procedure. The fourth column indicates what range in effective
temperature the selected region shows strong sensitivity to varying parameters
and the fifth column indicates whether the region is more sensitive to 
effective temperature or surface gravity.}
\tablenotetext{a}{\citet{tor93}}
\tablenotetext{b}{\citet{kir91}}
\end{deluxetable}%

\begin{figure}[t]
\begin{center}
\includegraphics[width=3.25in]{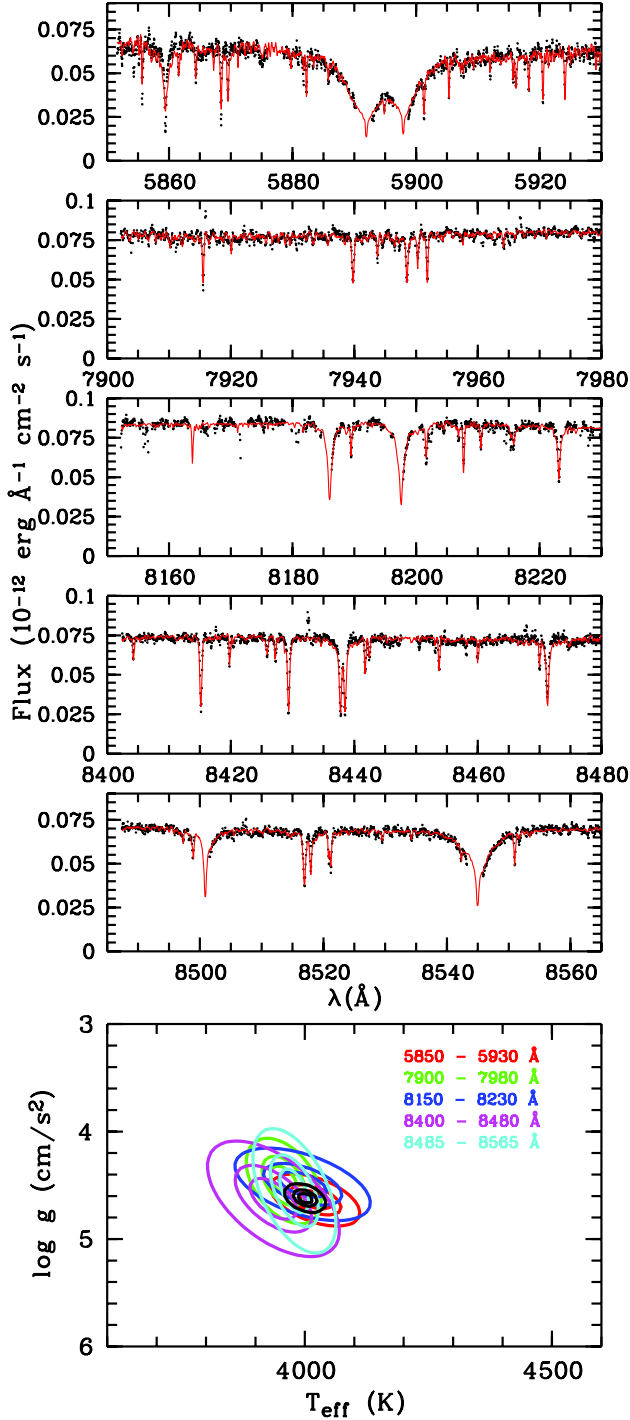}
\caption{Spectral fits (solid red lines) to $\eta$\,Cha~10
  (black dots), a K7 star, in the regions listed in Table
  \ref{tab:spectralregions}. Quality of fit ($\chi^2$) contours at 68.3,\%, 95.4\,\% and 99.99\,\%
  confidence for each spectral window are  shown in the bottom panel with each 
  level given by $\chi^{2}_{\mathrm{min}} \times [1 + (2.30, 6.17, 18.4)/N_{\rm
dof}]$, where $N_{\rm dof}$ is the degrees of freedom in the fit and
$\chi^{2}_{\mathrm{min}}$ is the best-fit value. The summed $\chi^2$ values for all 
spectral windows are represented by the tight solid black contours.  \label{fig:specfit1138}}
\end{center}
\end{figure}

\begin{figure}[t]
\begin{center}
\includegraphics[width=3.25in]{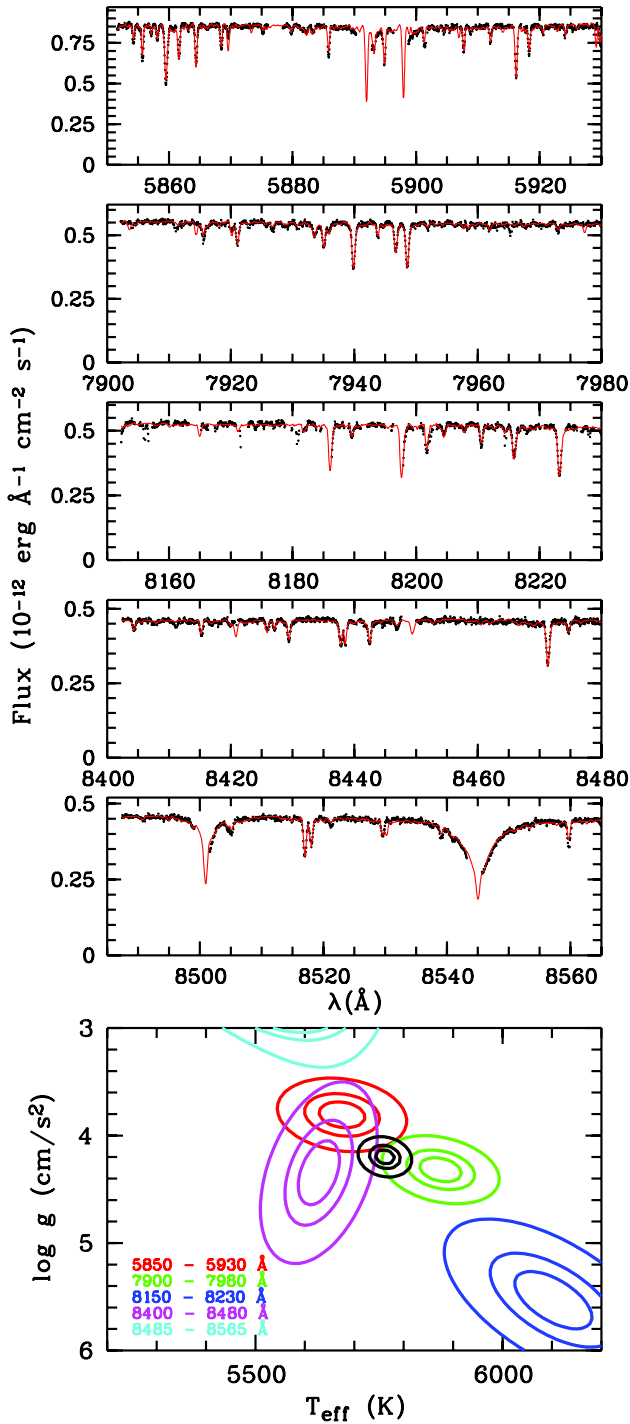}
\caption{As Fig.~1, but for HIP 32235 (black dots), a G6V-type TUCHOR
  member.   \label{fig:specfit8095}}
\end{center}
\end{figure}

In general, the dependence of these features on our parameters is
somewhat degenerate, as different $T_{\mathrm{eff}}$ and $\log g$
combinations can fit a given absorption feature equally well. As
discussed in \citet{moh04}, features that show contrasting dependence
on gravity and temperature are needed to constrain the parameter
space.  Thus, we choose combinations of molecular absorption features
like titanium oxide (TiO) and vanadium oxide (VO) with lines from
neutral alkali elements, like \ion{K}{1}, \ion{Na}{1} and
\ion{Mg}{1}. This works well because, for neutral alkali elements, an
increasing temperature can be compensated by increasing gravity, while
the molecular absorption bands show less correlation between gravity
and temperature. For instance, TiO bands in the vicinity of
7\,100\,\AA\ are much more sensitive to temperature than to gravity,
while the triple headed TiO bands in the range of 8\,440\,\AA\ are
more sensitive to gravity than to temperature. Examples of this
dependence can be seen in the $\chi^2$ contours shown in the bottom 
panels of Figs.~\ref{fig:specfit1138} and \ref{fig:specfit8095}.  Of course, 
these dependencies change with temperature, and hence different sets of
regions are best for different spectral types. In principle, one could
choose to fit different regions in different temperature ranges, but

this risks (borne out in practise) that at the borders there are false
jumps in parameters due to systematic problems with the models. Thus,
we fit all regions for all stars, which gives very tight constraints
on $T_{\mathrm{eff}}$ and $\log g$ for all spectral types (and more
smoothly varying systematic offsets; see below).  After investigating
spectral fits of twelve individual regions collected from
\citet{tor93} and \citet{kir91}, we settled on the five separate
spectral regions listed in Table \ref{tab:spectralregions}.  Also
listed is the effective temperature range where the features provide
strong constraints, as well as whether these constraints are
predominantly on $T_{\mathrm{eff}}$ and/or $\log g$.

We excluded regions which either tended to overestimate $\log g$ or
provided $T_{\mathrm{eff}}$/$\log g$ values extremely offset from the
majority of spectral regions. For instance, for many of the neutral metal
lines, particularly \ion{K}{1} at 7\,665 and 7\,699\,\AA, the best fits are 
at surface gravities much higher than are expected for
late-type PMS stars. Since these lines are sensitive to gravity and generally fit
the spectrum very well, the entire fit was very sensitive to the
\ion{K}{1} line. Due to this inconsistency between the line strengths
in the atmospheric models and predicted surface gravity, this region
and several others were not used. We note that we 
use other neutral metal lines such as \ion{Na}{1} that could be flawed 
similarly. In practise though, these produced fits that were 
consistent with the other spectral regions investigated.

\subsection{Fitting Methods}\label{s:methods}

All spectra in our study with spectral types later than F5
(corresponding to about 6\,500\,K, the highest temperature for which
we have models) were fitted to the synthetic spectra for each spectral
window listed in Table \ref{tab:spectralregions}. Prior to the fit,
the synthetic spectra were convolved with a gaussian filter to match
the observed resolution. The fit used a variant of the
broadening-function formalism introduced by \citet{ruc02}. In this
formalism, a least-squares fit is made of the observed spectrum to a
set of reference spectra that differ only in their velocity offset.
This way, rotational broadening is automatically accounted for. The
difference with the formalism of \citeauthor{ruc02} is that the sum of
the model spectra at various velocities is also multiplied with a
polynomial, to account for not only the normalization, but also
for small errors in the flux calibration. For our small wavelength
regions, we found that a third-degree polynomial sufficed.

We note that our fitting method effectively introduces a relatively
large number of parameters that are not of direct physical interest,
viz., the line shape. Since these parameters might be covariant with
some of the effects of temperature and/or gravity changes, the
constraints on $T_{\rm eff}$ and $\log g$ we find are thus not as
strong as might be possible if, e.g., instead we broadened the
synthetic spectra with an analytic broadening model (with only the
projected rotational velocity as parameter). In practice, however,
the covariance is small, and our errors are clearly dominated by
systematic mismatches between the models and the observations.

With the above least-squares fitting method, we fitted each of the
five regions from each observed spectrum to 77 of the synthetic
spectra, covering all seven values of $\log g$ (from 3.0 to 6.0 in
steps of 0.5), and eleven temperatures in a range of 1\,000\,K around
the effective temperature inferred from the object's spectral type (as listed
in Tables~\ref{tab:TWA}--\ref{tab:RVstd}; we used the temperature scale
for dwarfs of \citealt{luh98}). For new companions with unknown
spectral types, initial estimates for the temperatures were found
through trial and error. 

To determine combined constraints on $T_\mathrm{eff}$ and $\log g$ for
a given spectrum, we add up the $\chi^{2}$ values derived for the five
regions. The results are similar to fitting the five regions
together, except that our method leaves greater freedom for variations
in continuum (which might be expected) or line shape (which would
not). The difference with fitting the entire spectrum is that we have
effectively ignored the parts of the spectrum that are either
insensitive to $T_\mathrm{eff}$ and $\log g$ or where the models are
shown to fit poorly to the observations. 

For our final estimates of the best-fit values of $T_\mathrm{eff}$ and
$\log g$ for each spectrum, we interpolate in the 7 by 11 grid by
determining the minimum of a two-dimensional parabola (of the form $a
+ bx + cy + dx^{2} + exy + fy^{2}$) fit to the sixteen grid points
with the lowest values of $\chi^{2}$.

\subsection{Examples}\label{s:example}
 
In Fig.~\ref{fig:specfit1138} and \ref{fig:specfit8095}, we show our
fits for one spectrum of each of the stars $\eta$\,Cha~10 and TUCHOR member
HIP\,33235. The gaps in the data in the panels centered on 5\,890
and 8\,525\,\AA\ are due to the removal of emission lines as discussed
in \S\ref{s:mods}. Similarly, on close inspection, one sees that
multiple telluric lines have been removed from the 8\,190\,\AA\
region.

Overall, the broadened models reproduce the observed spectra well. In
detail, however, there are clear inconsistencies between the observed
and synthetic spectra, apparent in many of the panels. In particular,
a number of lines appear to be absent in the models, or are clearly
too weak. We also found examples of the reverse in some other
wavelength regions.

In the bottom panels of Figs.~\ref{fig:specfit1138} and
\ref{fig:specfit8095}, the resulting 68.3\,\%, 95.4\,\% and 99.99\,\% confidence contours in the
$T_{\mathrm{eff}}$-$\log g$ parameter space are shown, with levels set
according to \citet{pre92} at $\chi^{2}_{\mathrm{min}} \times [1 + (2.30, 6.17, 18.4)/N_{\rm
dof}]$, where $N_{\rm dof}$ is the degrees of freedom in the fit and
$\chi^{2}_{\mathrm{min}}$ is the best-fit value inferred from the
parabolic fit. Contours are shown for each spectral window as well as
for the total $\chi^{2}$ values. For the different
spectral regions, one notices the different covariances in
$T_{\mathrm{eff}}$ and $\log g$, and how, by using a number of these
sensitive regions, it is possible to constrain gravity and temperature
precisely. Indeed, the contours for the summed $\chi^{2}$
distribution (solid black contours) are extremely tight. We will see below that different
spectra of the same star lead to similarly small scatter in the
inferred temperature and gravity. 

One also notices, in particular in Fig.~\ref{fig:specfit8095}, that the
contours for the different regions are statistically inconsistent with
each other, with differences of several 100\,K in
$T_{\mathrm{eff}}$ and up to 1\,dex in $\log g$.  These differences
likely reflect systematic uncertainties in the models, similar to what
we find for the resulting best-fit average values below.

\begin{figure}[t]
\begin{center}
\plotone{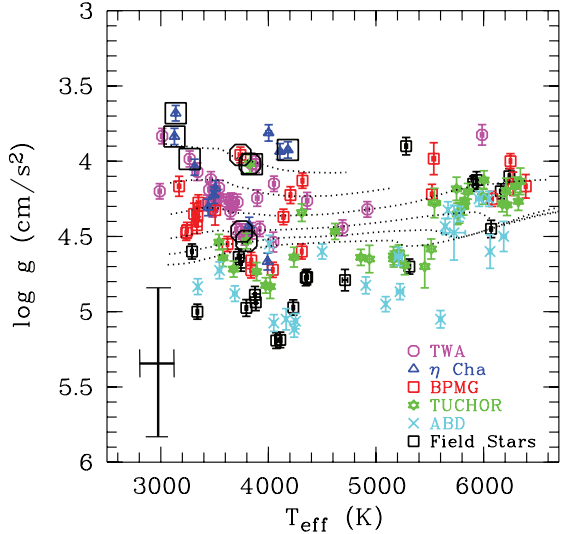}
\caption{Surface gravity, $\log g$, as a function of effective
  temperature, $T_{\mathrm{eff}}$, for all five PMS groups, as well as
  field stars. PMS tracks from \citetalias{bar98} are overdrawn, with, 
  from top to bottom, ages of 4, 8, 15, 25, 35, 45~and~150\,Myr. 
  We represent the external error in the plotted parameters 
  for the ensemble in the bottom-left corner. It is evident that there are
  systematic problems with the values of $\log g$ derived from the
  models, particularly for $3\,500< T_{\mathrm{eff}}<4\,100\,$K.
  However, the relative ordering is clear: stars that are older, such
  as ABD and TUCHOR members, have higher surface gravities than
  younger stars. The ultrafast rotators and accretors in
  our sample are marked with circles and squares, respectively. \label{fig:logg_vs_t}}
\end{center}
\end{figure}

\subsection{Results}
\label{s:LiRes}

The best-fit model $T_{\mathrm{eff}}$ and $\log
g$ with corresponding errors, for each star, including field stars,
can be found in Tables~\ref{tab:TWA}--\ref{tab:RVstd}. We consider two
ways of estimating the associated uncertainties. First, we follow
\citet{pre92} and use the curvature of the best-fit two-dimensional
parabola to the $\chi^2$ values to find regions which are enclosed
within a level of 68\% confidence.  Second, we consider the error in
the mean between the results from different spectra of the same source
taken at different epochs (e.g.\ the standard deviation divided by the
square root of the number of observations taken). 

For the temperatures, our statistical uncertainties are typically around 9\,K,
 while the scatter derived from multiple observations
of the same object is on average 11\,K.  For the surface gravities, our 
statistical uncertainties numbers are 0.02\,dex on average, and the 
scatter derived from the error in the mean from multiple observations 
is about 0.03\,dex. The above suggests the true intrinsic uncertainties 
($\sigma_\mathrm{int}$) in our temperature
and gravity measurements are very small, about 10\,K and 0.05\,dex
for a single observation.  We will see below, however, that systematic
mismatches as a function of, e.g., spectral type, are much larger.

In Fig.~\ref{fig:logg_vs_t}, we show the distribution of $\log g$ as a
function of $T_{\mathrm{eff}}$ for each group as well as for the field
stars observed. In addition, we draw isochrones from the
\citetalias{bar98} PMS evolutionary models.  One immediately sees
there is a clear systematic problem in determining $\log g$ for stars
with $3\,500<T_{\mathrm{eff}}<4\,100$\,K: $\log g$ increases
 with temperature from 3\,000\,K to 4\,000\,K, but at around 4\,000\,K, 
 it becomes almost 1 dex smaller, an unrealistic
 physical trend. The systematic differences in $\log g$ can result from a few
effects. If the resolution of the model spectra is not much better
than the observations before smoothing, line
depths can be systematically off.  In addition, stellar activity also
introduces systematic errors, as has been found for young M-dwarfs,
where the chromosphere feeds back into the photosphere \citep{fuh05}.
This feedback is not incorporated into the model, and may lead to
systematic errors of about 0.5 dex in $\log g$.

To evaluate the dependence of the fitted $T_\mathrm{eff}$ on $\log g$, we re-fit 
our spectra three more times using the same method described in 
\S\ref{s:mods}, except with $\log g$ fixed to 4.0, 4.5 and 5.0. The change 
in $T_\mathrm{eff}$ across the surface gravity space is small in an absolute 
sense. Going from log g fixed at 4.0 to 4.5, the average change in temperature
within our sample is $\Delta T_{\mathrm{eff}} = 71$\,K. It is 
slightly higher going from 4.5 to 5.0, with
$\Delta T{\mathrm{eff}} = 124$\,K. Changes from $\log g =$ 4.0 to 5.0, yield absolute 
average changes in $T_\mathrm{eff}$ of 162 K.

Thus, the external errors related to the models are much higher than 
the internal errors. Tables 2 to 7 list the best-fit surface 
gravities and temperatures with quoted errors representing the error 
in the mean of multi-epoch observations of the object. More conservatively,
we consider our external errors to be 150 K in $T_\mathrm{eff}$ and 
 0.5 dex in $\log g$.

\begin{figure}[t]
\begin{center}
\plotone{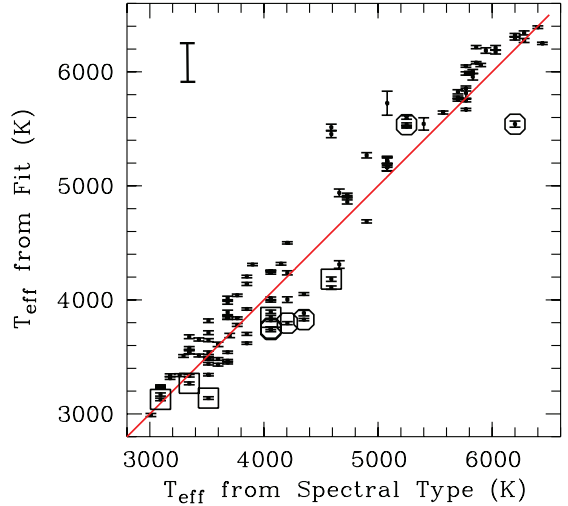}
\caption{Comparison of the temperatures obtained from spectral fitting
  (see \S\ref{s:methods}) with those derived from spectral types using
  the scale of \citet{luh98}. We represent the external error in the plotted parameters for the ensemble in the top left corner. The ultrafast rotators and accretors in
  our sample are marked with circles and squares, respectively. \label{fig:fitt_vs_sptt}}
\end{center}
\end{figure}

In Fig.~\ref{fig:fitt_vs_sptt}, we compare our computed temperatures
to those obtained by converting spectral types to effective
temperatures using the temperature scale for dwarf-type stars of
\citet{luh98}. Generally, the two scales are consistent within the 
external errors just discussed, with the largest deviations
show by the accretors (marked by squares) and ultrafast 
rotators (marked by circles, see also \S\ref{s:rotation}). 
In addition, most of the objects with
$T_\mathrm{eff} >5000$\,K, while individually consistent within the
uncertainties, appear systematically to have fitted temperatures
slightly greater than those inferred from spectral type, suggesting a
small systematic error.

\section{Lithium Abundances}\label{s:Li}
 
We analyse lithium abundances using two independent methods. In one
method, which follows previous work \citep{jef05}, we measure the
equivalent width (EW) of the 6\,708\,\AA\ lithium absorption doublet
for each spectrum. We use this measurement and the spectral types quoted
in the literature to chronologically order the groups. The other
method uses the best-fit model calculated in \S\ref{s:gravtemp} to fit
model spectra for various lithium abundances to regions surrounding
the 6\,104 and 6\,708\,\AA\ lithium absorption features. As with our
EW measurements, we chronologically order the groups. Further, we 
compare the observations with PMS models of \citetalias{bar98} 
and \citet{sie00} to see if the isochrone ages based on lithium 
abundances are consistent with other age determinations 
of the five nearby PMS groups in this study

\subsection{Lithium equivalent widths}\label{s:ew}

First, we use the EW of the 6\,708\,\AA\ lithium doublet feature in
each observed spectrum as an empirical measurement
of the lithium abundance. To measure it, we interactively chose
the edges of the lithium feature and the boundaries of sufficiently 
large regions around it to define the stellar continuum. We chose 
to consistently reject the lowest 15\%
of the flux points in the continuum region (corresponding to the
deepest absorption features), so that we do not underestimate the
continuum flux. We do not apply any 
correction for the 5 identified accretors in our sample, but find 
that veiling due to accretion can reduce the measured EW, 
as discussed briefly below.

\begin{figure}[t]
\begin{center}
\plotone{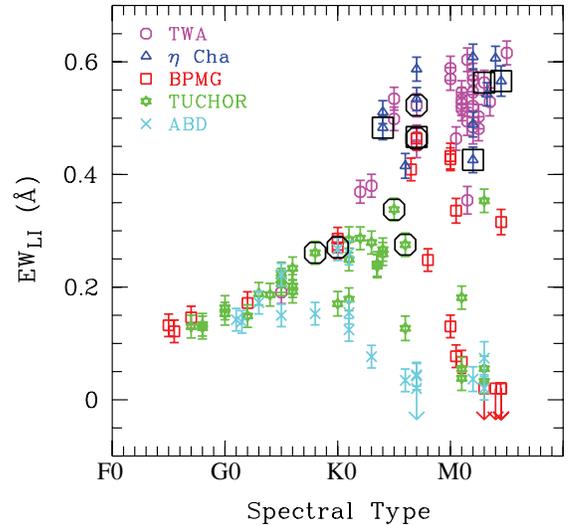}
\caption{Equivalent width (EW) of the 6708\,\AA\ lithium line as a
  function of spectral type. Without depletion, the EW would increase
  with decreasing temperature, as is seen with the youngest two
  groups, TWA and $\eta$\,Cha.  But in older groups lithium is being
  depleted, and the order, from youngest to oldest, is evident: BPMG, TUCHOR,
  and ABD.  We identify the ultrafast rotators and accretors in our
  sample with black circles and squares, respectively. \label{fig:ew_vs_t}}
\end{center}
\end{figure}

The resulting mean EWs, averaged over all available spectra, are
listed in Tables~\ref{tab:TWA}~to~\ref{tab:ABD}, with uncertainties
being the error in the mean between multiple spectra.
Fig.~\ref{fig:ew_vs_t} displays the resulting EWs as a function of
spectral type. From this purely empirical figure,
the chronological order of the groups is evident: from oldest to
youngest, they are ABD, TUCHOR, BPMG, TWA and $\eta$\,Cha. 
The same ordering was found by \citet{zuc04} from lithium EWs for
a smaller sample of stars. Although
some TWA members appear to be as young as those in $\eta$\,Cha, it is
quite clear that no members are older than BPMG, contrary to the
suggestion by \citet{law05}. (Note that we implicitly assume 
here the initial lithium abundance was
the same for all groups.  We return to this below.)

We identify the ultrafast rotators and accretors in our sample with
black circles and squares, respectively. The EWs of the rotators in
BPMG (red squares) and TUCHOR (green stars) stand out above the
general trend for each group; we will return to this in
\S\ref{s:rotation}.  We also note that one accretor in $\eta$\,Cha,
($\eta$\,Cha\,13) has a lithium EW that is lower than most objects at
the same temperature ($\sim3100\,$K) and age, but we will find below a
lithium abundance near initial from the fits to the spectra of this
star (see \S\ref{s:Li-method}).  This suggests the low estimate of the
EW is related to the fact that it is accreting, either by its effect
on the continuum level or alternatively, by us 
underestimating the temperature.  Indeed, the
latter may well play a role: from our spectral fit, we infer
$T_\mathrm{eff}=3139\pm96$\,K, while that from its M2 spectral type,
one would estimate $T_\mathrm{eff}=3514$\,K \citep{luh98}.  With the latter
temperature, the object would match the trend much better. Because of the 
large accretion lines in the spectra, the models are a poor representation
of the data throughout the regions selected for our fits. A higher 
temperature model is just a slightly worse fit to the observations.

\subsection{Lithium Line Analysis}\label{s:Li-method}

\begin{figure*}[t]
\begin{center}
\plotone{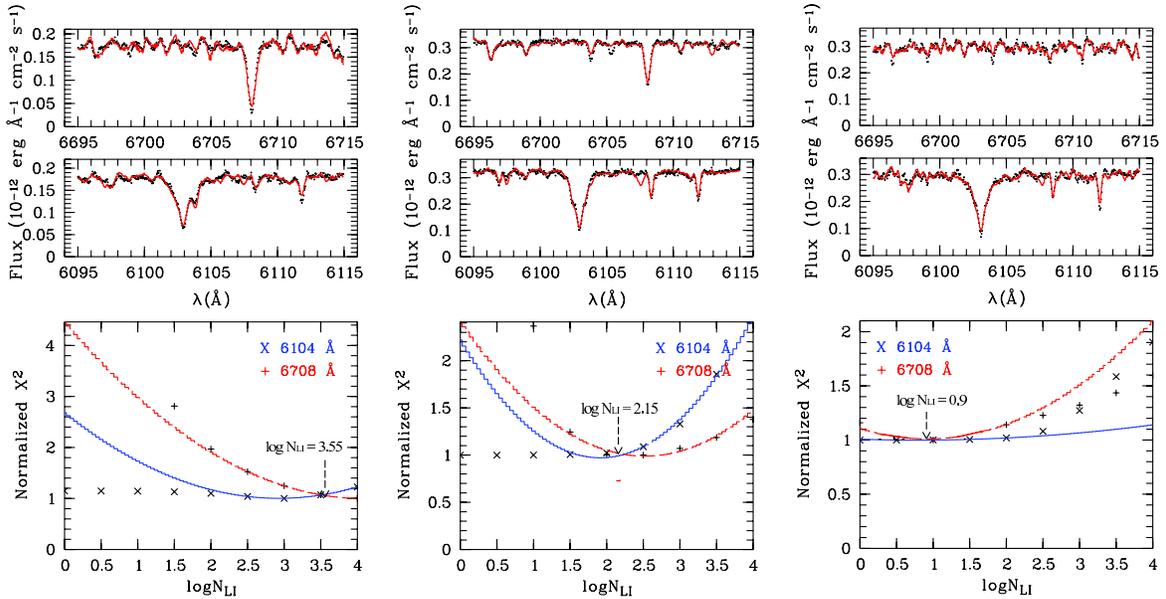}
\caption{Examples of fitting the two lithium features to three stars
  (from left to right: TWA 02A (M2e), CD-60\,416 in TUCHOR
  (K3/4) and GJ 3305 in BPMG (M0.5)).  For each star, the top panel
  shows the fit to the strong 6\,708\,\AA\ lithium doublet and the
  middle panel the fit to the much weaker 6\,104\,\AA\ lithium line.
  The bottom panel shows the normalized $\chi^{2}$ values for the
  abundances at which we had models, and the parabolic interpolations
  used to determine the best-fit abundances (which is taken to be at
  the minimum of the sum of these two curves; see
  \S\ref{s:Li-method}). \label{fig:lifit-montage} }
\end{center}
\end{figure*}

Using the average best-fit effective temperatures and surface
gravities listed in Tables~\ref{tab:TWA}~to~\ref{tab:RVstd}, synthetic
spectra, varying in lithium abundance from $\log N_{\mathrm{Li}}$ =
0.0 to 4.0 in steps of 0.5, are, for each spectrum, fitted to small
spectral regions around the 6\,104\,\AA\ and 6\,708\,\AA\ lithium
absorption lines. The fitting method is identical to the method
outlined in \S\ref{s:methods}. However, we constrain the minimization
of the least-squares fit in a slightly different manner.

Synthetic spectra of varying lithium abundance are fitted to
20\,\AA\ wide spectral regions -- specifically
6\,095--6\,115\,\AA\ and 6\,695--6\,715\,\AA.  The
6\,104\,\AA\ lithium triplet line is a weaker transition than the
6\,708\,\AA\ doublet line. It is also blended into the strong
6\,103\,\AA\ \ion{Ca}{2} absorption line. As a result, the line is
only detectable for high lithium abundance. Overall, we are not very
sensitive to this line, detecting only a small change in $\chi^{2}$
over the entire lithium abundance range. This is not the case with the
stronger lithium doublet at 6\,708\,\AA, which is very sensitive to
lithium abundance, and shows sharp transitions from good to bad in its
least-squares fits.  For high lithium abundances, however, the line
saturates and without the 6\,104\,\AA\ line no good abundance
estimates are possible.

In order to be able to treat all data uniformly, irrespective of
lithium abundance, we proceeded as follows.  First, we use the
6\,708\,\AA\ region, with its higher sensitivity, to determine the
approximate abundance, and select the points with $\log N_\mathrm{Li}$
corresponding to the 4 lowest $\chi^{2}$ values from the fit to this
region.  For both spectral regions, we then fit a 2nd order polynomial
to the $\chi^2$ for these 4 selected $\log N_\mathrm{Li}$.  Next, we
normalize both fitted polynomials by dividing by the minimum
$\chi^{2}$ value for each spectral region.  The two normalized curves
are then added together to give an average curve, and the minimum of
this curve is what we take to be the best-fit $\log N_\mathrm{Li}$.
Thus, in our procedure for determining the lithium abundances, 
we give equal weight to both regions, unlike
the procedure outlined in \S\ref{s:methods}, where we desired to keep
the weight on the best fitted regions by summing the raw $\chi^{2}$
values.

By way of example, we show, in Fig.~\ref{fig:lifit-montage}, the
spectral fits to the two lithium features for three stars with
distinctly different lithium abundances.  The resulting normalized
$\chi^2$ curves are shown in the bottom panel for each star.  Averages
of the best-fit $\log N_\mathrm{Li}$ for all stars can be found in
Tables~\ref{tab:TWA}~to~\ref{tab:RVstd}, with uncertainties
representing the error in the mean between multi-epoch observations.

As discussed in \S\ref{s:LiRes}, uncertainties in the model atmospheres
lead to larger errors than the internal errors quoted in 
Tables~\ref{tab:TWA}~to~\ref{tab:RVstd}. Thus, we also handle external 
errors for the lithium abundances with the same approach. We 
investigate changes in the fitted abundances by re-fitting the 
lithium lines to models with $T_{\mathrm{eff}}$ perturbed by $\pm 100$
from the initial best-fit $T_{\mathrm{eff}}$ and also to models with
$\log g$ also perturbed by $\pm 0.5$\,dex. We find that this 
parameters leads to changes in  $\log N_\mathrm{Li}$ of 0.15 on average. 
We show this error on all relevant figures to indicate our estimate of the
external uncertainties.

\begin{figure}[ht]
\begin{center}
\plotone{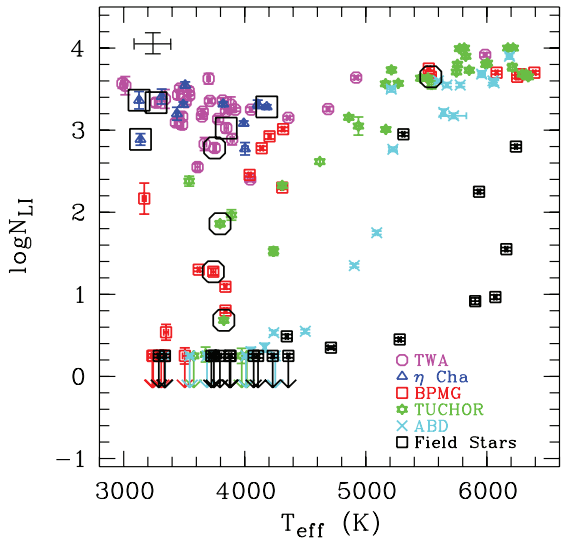}
\caption{Lithium abundance derived by fitting the 6\,104\,\AA\ and
  6\,708\,\AA\ spectral features as a function of temperature. We 
  identify the ultrafast rotators and accretors in our
  sample with black circles and squares, respectively.\label{fig:li-vs-t}}
\end{center}
\end{figure}

\subsection{Ages from the lithium abundance versus temperature isochrones}\label{s:ages}

As a PMS group ages, the lithium abundances of group members deplete
as a function of luminosity and time (see \S\ref{s:intro}). As with EWs in \S\ref{s:ew}, 
we can order the groups in age using the relative depletion of
members of different groups and by comparison with models, 
we can also determine absolute ages.  Of course, the
absolute ages will only be as good as the models. To get an idea of 
the associated uncertainty, we try both the models of \citetalias{bar98}
and \citet{sie00}.

In Fig.~\ref{fig:li-vs-t}, we show the distribution of measured
lithium abundance as a function of temperature for each of the groups
in this study, as well as for the field stars (which should be fully
depleted).  As with the EWs in \S\ref{s:ew}, it is easy to order the groups
chronologically based on the lithium depletion distribution: from
oldest to youngest, we again find the order ABD, TUCHOR, BPMG, and
then $\eta$\,Cha and TWA both at about the same age. It is also clear
that the field stars are older than all of the PMS groups in this
study.

\begin{figure*}[t]
\begin{center}
\includegraphics[width=6.5in]{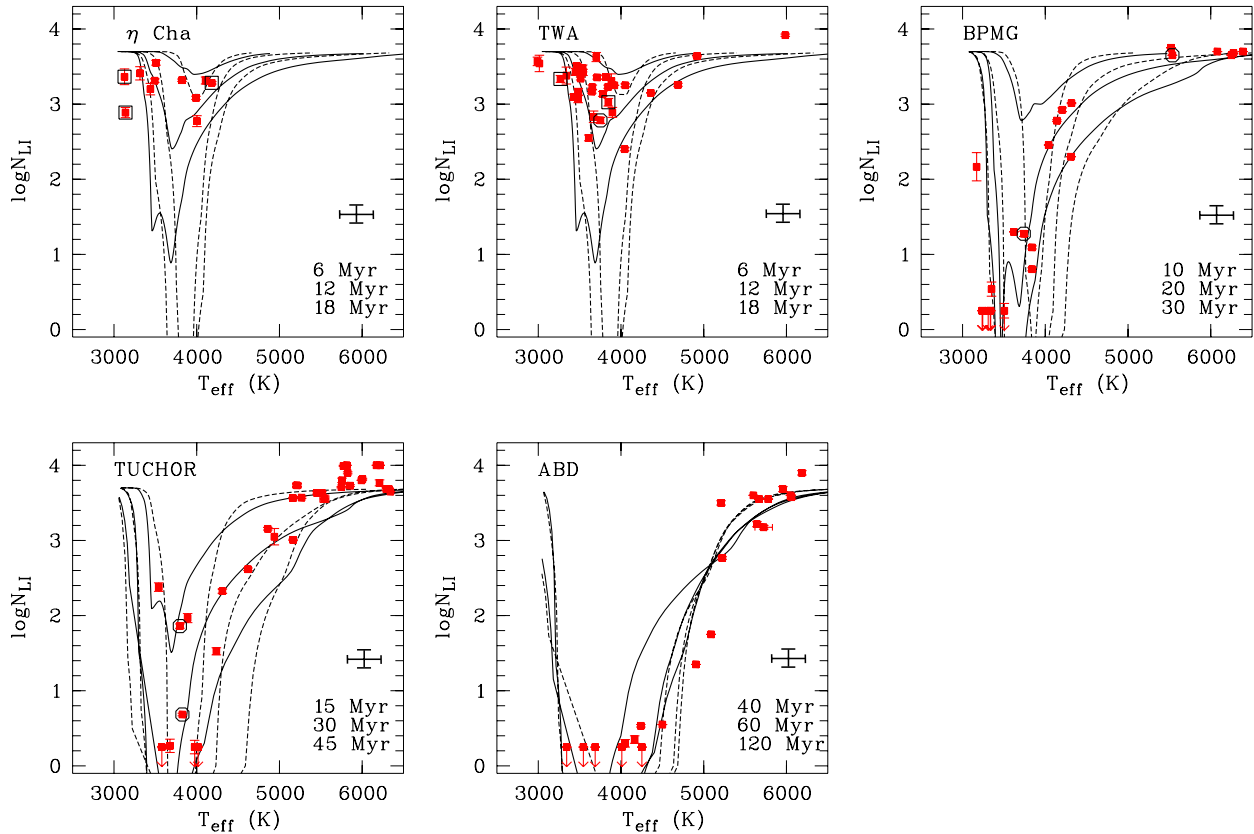}
\caption{Lithium abundances derived from fits to the spectra as a
  function of temperature for each group.  Overdrawn are predictions
  from the evolutionary models of \citetalias{bar98} (solid line) and 
  \citet{sie00} (dashed line) as indicated by the ages in 
  the bottom-right corner of each plots. Above the ages, 
  external errors are shown. In addition, the ultrafast rotators are
  identified by black circle outlines, and accretors are identified
  by black square outlines.  \label{fig:li-vs-t-montage}}
\end{center}
\end{figure*}

In Fig.~\ref{fig:li-vs-t-montage}, the dependence of $\log
N_{\mathrm{Li}}$ on stellar $T_{\mathrm{eff}}$ is shown for each group
individually, and a range of PMS isochrones from \citetalias{bar98} 
(solid lines) and \citet{sie00} (dashed lines).  Both models use a convection
mixing length of $\alpha_{\mathrm{MLT}}=1.9$  and are scaled to the initial lithium 
abundance. We chose an initial lithium abundance of  
$\log N_{\mathrm{Li}} = 3.7$ to match the abundance we measured
from our method (see \S\ref{s:Li-method}) for the majority of undepleted
 stars in the entire sample. The inferred ages do depend on this
 choice, as well as the choice of model.
 
 From Fig.~\ref{fig:li-vs-t-montage}, we can infer ages by eye for each group.
 We tried but decided not to use quantitative analysis, because it is 
 apparent that the models do not reproduce the data to enough accuracy,
 especially in the temperature range where the lithium abundances are 
 most dependent on age. 
In general, we find that the models of \citetalias{bar98}
 fit the nature of the depletion slightly better for all of the groups, 
 but that the ages derived are similar, apart from a small
systematic offset (with \citetalias{bar98} giving slightly higher ages).

  We find that a lithium depletion age of $12{\pm6}$\,Myr for $\eta$\,Cha 
 using the models of  \citetalias{bar98} ($12{\pm8}$\,Myr using the 
 \citet{sie00} models; hereafter given in parentheses), and and 
 an age of $12{\pm8}$\,Myr ($12{\pm8}$\,Myr) for TWA,
consistent with dynamical expansion ages \citep{jil05,del06} and other
age estimates also based on \citetalias{bar98} PMS models 
\citep{luh04,zuc04,bar06}.  For BPMG, we find an age of
$21{\pm9}$\,Myr ($13{\pm5}$\,Myr), which is in agreement with the estimate of
9--17\,Myr from other methods \citep{zuc04,fei06}.  It is slightly
higher than its dynamical expansion age of 11.5\,Myr from
\citet{ort02}, but agrees with a different age based on lithium dating
of 10--20\,Myr recently found by \citet{mam07}.  For TUCHOR, we find
an age of $27{\pm11}$\,Myr ($22{\pm10}$\,Myr), which is consistent with
all previous age estimates for this group \citep{zuc00,ste00}. 

For ABD, we find that it is clearly older than TUCHOR and
clearly younger than the field stars, however the age estimate from
PMS models is poorly constrained.  Although the field stars show
more depletion than ABD, this is not predicted by the PMS isochrones;
both models show no depletion after $\sim\!45$\,Myr for stars
with $4\,000<T_{\mathrm{eff}}<6\,000$\,K. Until the PMS evolutionary
models are improved, the best way to find an
upper limit to the age of the ABD group would be to use the cool end of the LDB.
This would require stars with spectral types later than M3
($T_{\mathrm{eff}}\lesssim$3\,300\,K), but, unfortunately, no such
members are known in ABD.  Within our present large uncertainties, our
age estimate is consistent with both a younger estimate of 50\,Myr
based on H$\alpha$ emission strength \citep{zuc04b}, as well as an
older one of 100--140\,Myr from HR isochrones \citep{luh05} and
dynamical expansion \citep{ort07}.

We close with a number of notes.  First, while the poor fit of the
data to the models leads to rather large uncertainties on the ages,
these should be considered overall shifts: the age ordering of the
groups is secure.  Second, an additional uncertainty in 
the derived ages is our choice of initial
lithium abundance, of $\log N_{\mathrm{Li}}=3.7$ based
on our observations. Decreasing the initial lithium abundance to
$\log N_{\mathrm{Li}}=3.3$ (as used in \citet{jef05}), predicts 
younger group ages by about 5 Myrs. On the other hand, using a higher
initial lithium abundance of $\log N_{\mathrm{Li}} = 4.0$, yields
ages larger by 5-10 Myrs. Third, we have ignored non-LTE effects in the lithium lines
(\S\ref{s:models}).  While our scatter is larger than the predicted
effects, the systematic changes with temperature and abundance will
lead to additional systematic age differences.  
It also may be the underlying reason for our need for
a relatively high initial abundance: \citet{car94} found that around
6000\,K, the correction for the 6708 line is about $-0.3\,$dex, which
would imply initial abundances more in line with expectations (at
these temperatures, the 6\,104\,\AA\ line is very weak and contributes little to our
fits.  

\subsection{The Effect of Rotation on Lithium Depletion}\label{s:rotation}

We examine the effect of stellar rotation on lithium depletion using 
projected rotational velocities found previously from our observations
\citep{jay06,sch07}. In Fig.~\ref{fig:li-vs-t-montage}, we identify 
ultrafast rotators as those stars that have $v$\,$\sin i>70$\,km\,s$^{-1}$.
One member (TWA 6) from TWA, two members (PZ\,Tel, CD\,64d1208) of BPMG and 
three members (CD\,53\,544, HIP\,108422, HIP\,2729) of TUCHOR are identified
as ultrafast rotators. In three of the cooler stars ($T_\mathrm{eff}\sim$3800\,K), CD\,64d1208, 
CD\,53\,544 and HIP\,2729 , the lithium EWs are noticeably higher, 
and the derived abundances larger, than the trend in lithium depletion for 
the entire group. The correlation between fast rotation and slower lithium 
depletion has also been seen previously in a sample of weak-line T Tauri stars 
\citep{mart94} and in the 115\,Myr Pleiades cluster \citep{sod93,gar94}. 
It may be related to rapidly rotating stars being relatively cooler as the rapid 
rotation inhibits convection \citep{cha07}. This alternative view is supported by the location of the rotators in Fig.~\ref{fig:fitt_vs_sptt}. It is evident that the $T_\mathrm{eff}$ 
derived for these rotators is 100--300\,K cooler than the temperature 
derived from their spectral types. It may be that the presence of a colder 
equatorial region and a hotter polar one affects the model fits differently than
the spectral typing.

However, the trend that lithium depletion is slowed down by rotation
is not seen in all of our ultrafast rotators. The relatively slower rotators 
PZ\,Tel ($v$\,$\sin i=77.5$\,km\,s$^{-1}$) and TWA\,6 ($v$\,$\sin i=79.5$\,km\,s$^{-1}$)
have lithium abundances comparable to other members in their groups. 
For our faster rotator, HIP\,108422 ($v$\,$\sin i=$139.8\,km\,s$^{-1}$),
the spectrum is so strongly broadened that our fitting method
does a poor job and we see no change in the quality of fit
for varying lithium abundances. For this, reason, we do not quote an abundance
for this object, but do note, however, that the equivalent width measured is consistent with 
other group members.

\subsection{Notes on Individual Systems}

\textit{BD\,17$\,^{\circ}$6128 --} This binary system from BPMG
consists of a K7 primary with $T_{\mathrm{eff}}=$4\,140\,K and 
$\log N_{\mathrm{Li}}=2.78$ and a lithium depleted secondary of 
$T_\mathrm{eff}=$3\,350\,K and $\log N_{\mathrm{Li}}=0.54$. This
system is unique to our sample as it is the only case of the cool end
of the LDB in effect within a binary, and provides a precise, if model 
dependent, age of the system. Using the models from \citetalias{bar98}, an age of
15--50\,Myr is predicted, consistent with the age we inferred for BPMG
as a group. 

 \textit{GSC\,08056-0482 --} The measured $\log N_{\mathrm{Li}}$ for TUCHOR member
GSC\,08056-0482 is much higher than expected for a $\sim\!30$\,Myr
old, M3 dwarf.  Indeed, another M3 star in the younger BPMG,
GSC\,08491-1194, has an abundance over two orders of magnitude smaller
(consistent with models at $\sim\!20\,$Myr). With a modest $v\sin i$
of $34.2{\rm\,km\,s^{-1}}$ \citep{jay06}, rotation does not explain
the high lithium abundance. As pointed out already by previous authors
(see Table 3 of \citealt{zuc04}), the lithium abundance suggests that
GSC\,08056-0482 is likely not a member of TUCHOR, but a star slightly
older than TWA and $\eta$\,Cha, but definitely younger than BPMG.

\section{Summary and outlook}\label{s:conclusions}

We have measured effective temperatures, surface gravities, lithium
equivalent widths and lithium abundances for 121 low-mass 
PMS stars from five nearby, PMS
groups ranging in age from 8--125\,Myr by performing least-squares
fits of high resolution spectra to synthetic spectra created from
PHOENIX model atmospheres \citep{hau98}. To investigate the
reliability of our measurements we compare the derived
$T_\mathrm{eff}$ and $\log g$ with isochrones from PMS evolutionary
models \citepalias{bar98} as well as temperatures derived from
spectral types. 

Isochrones from PMS models for $\log N_{\mathrm{Li}}$ as a function of
$T_\mathrm{eff}$ are visually compared to the observed distribution. 
We find agreement between ages derived from PMS isochrones of
\citetalias{bar98} and \citet{sie00} to ages calculated from other methods such
as dynamical expansion ages. We find that $\eta$\,Cha and TWA have
ages of $12{\pm6}$\,Myr and $12{\pm8}$\,Myr,
respectively. BPMG has an age of $21{\pm9}$\,Myr, and TUCHOR has
an age of $27{\pm11}$\,Myr. We can only constrain a tight lower
limit for ABD, with an age greater than $45$\,Myr, since, according
to the PMS models, there is no more lithium depletion after
$\sim\!45$\,Myr for stars with $4\,000<T_{\mathrm{eff}}<6\,000\,$K.
However, the halting of lithium depletion at this age and temperature
is inconsistent with observations of radial velocity standards which 
demonstrate more depletion than the ABD group and the model predictions.
Finally, we find that some of the ultrafast rotators in our sample have
significantly less lithium depletion than other stars in the same
group at the same temperature.

The consistent determination of $T_\mathrm{eff}$ and $\log g$ between
multiple epochs ($\sigma_T \simeq 10$\,K, $\sigma_{\log g} \simeq
0.05\,$dex) means that, in principle, we should be able to constrain
those parameters with this precision.  As revealed by
Fig.~\ref{fig:logg_vs_t}, however, there is an apparent
systematic offset between the $\log g$ inferred from model spectra and
$\log g$ expected from models of stellar evolution
(\S\ref{s:LiRes}). To account for these offsets, we introduce rough
conservative external errors by examining how the measured parameters
depend on each other. We find that the systematic errors in $\log g$ of 
0.5\,dex lead to systematic errors of 100\,K in our ability to constrain the 
$T_{\mathrm{eff}}$. These external errors, similarly lead to offsets in the 
measured lithium abundances of 0.15\,dex. 

The small internal errors that we have measured imply that currently our 
accuracy is limited by the models.  With further improvements in the atmospheric
models, there is a potential of comparing $T_\mathrm{eff}$ and $\log
g$ directly to evolutionary models, thereby finding an age constraint
independent of other estimators, such as color-magnitude diagram or
lithium depletion boundary fitting.  Our data set would be well-suited
for use with such future improved models.  Another use of our data set
would be to derive both overall metallicity and abundances for
individual elements.  While we do not believe this would affect our
derived temperatures, etc., to a significant degree, it may be
interesting to see how uniform the abundances are within (and between)
groups, and whether there is any dependence on binarity, etc.

\acknowledgements
We thank the referee for an excellent and thorough review, which
helped to improve our manuscript greatly.  We also thank the
outstanding support staff at Magellan for their assistance  
during multiple observing runs.  This work was supported in part by NSERC 
and the DFG (via Graduiertenkolleg 1351).  Some of the calculations presented 
here were performed at the H\"ochstleistungs Rechenzentrum Nord (HLRN); at the
NASA's Advanced Supercomputing Division's Project Columbia, at the
Hamburger Sternwarte Apple G5 and Delta Opteron clusters financially
supported by the DFG and the State of Hamburg; and at the National
Energy Research Supercomputer Center (NERSC), which is supported by
the Office of Science of the U.S. Department of Energy under Contract
No.\ DE-AC03-76SF00098. We thank all these institutions for a generous
allocation of computer time.

\clearpage

\begin{deluxetable}{llcllll}
\tablecaption{Results for stars in TW~Hydrae}
\tablecolumns{7}
\tablewidth{0pc}
%\tabletypesize{\small}
\tablehead{
	\colhead{Object ID} 			&
	\colhead{Sp.T.} 			&
	\colhead{\# Obs.}			&
	\colhead{EW$_{6708}$ (\AA)} &
	\colhead{$T_{\mathrm{eff}}$ (K)} 	&
	\colhead{$\log g$ }			&
	\colhead{$\log N_{\mathrm{Li}}$}		
}
\startdata
TWA 1          &   K7e\tablenotemark{a}  &  5 &  0.467 $\pm$   0.021 &   3847 $\pm$    16 & 4.02 $\pm$  0.07 &  3.03 $\pm$ 0.05\\
TWA 2A        &   M2e\tablenotemark{b}   &  5 &  0.567 $\pm$   0.021 &   3712 $\pm$    17 & 4.27 $\pm$  0.05 &  3.36 $\pm$ 0.03\\
TWA 3A        &   M3e\tablenotemark{b}   &  5 &  0.563 $\pm$   0.022 &   3268 $\pm$    12 & 3.98 $\pm$  0.05 &  3.34 $\pm$ 0.05\\   
TWA 3B        &   M3.5\tablenotemark{b}  &  5 &  0.529 $\pm$   0.025 &   3340 $\pm$    13 & 4.07 $\pm$  0.06 &  3.38 $\pm$ 0.11\\   
TWA 4AN       &                &  4 &  0.359 $\pm$   0.020 &   4045 $\pm$    24 & 4.54 $\pm$  0.07 &  2.40 $\pm$ 0.03\\   
TWA 4AS       &                &  4 &  0.442 $\pm$   0.020 &   4360 $\pm$    14 & 4.26 $\pm$  0.06 &  3.15 $\pm$ 0.02\\   
TWA 5A        &   M1.5\tablenotemark{b}  &  5 &  0.604 $\pm$   0.021 &   3432 $\pm$    12 & 4.28 $\pm$  0.05 &  3.09 $\pm$ 0.04\\   
TWA 6         &   K7\tablenotemark{a}    &  5 &  0.523 $\pm$   0.021 &   3751 $\pm$    12 & 4.47 $\pm$  0.05 &  2.78 $\pm$ 0.04\\   
TWA 7         &   M1\tablenotemark{a}   &  5 &  0.546 $\pm$   0.021 &   3541 $\pm$    10 & 4.18 $\pm$  0.05 &  3.42 $\pm$ 0.05\\   
TWA 8A        &   M2\tablenotemark{a}   &  5 &  0.546 $\pm$   0.021 &   3537 $\pm$    15 & 4.24 $\pm$  0.05 &  3.47 $\pm$ 0.05\\   
TWA 8B        &                &  5 &  0.575 $\pm$   0.026 &   3013 $\pm$    15 & 3.83 $\pm$  0.05 &  3.54 $\pm$ 0.11\\   
TWA 9A        &   K5\tablenotemark{a}   &  5 &  0.535 $\pm$   0.020 &   4052 $\pm$    12 & 4.15 $\pm$  0.05 &  3.25 $\pm$ 0.03\\   
TWA 9B        &   M1\tablenotemark{a}   &  5 &  0.517 $\pm$   0.021 &   3450 $\pm$    12 & 4.19 $\pm$  0.05 &  3.43 $\pm$ 0.05\\   
TWA 10         &   M2.5\tablenotemark{a} &  5 &  0.481 $\pm$   0.021 &   3512 $\pm$    12 & 4.24 $\pm$  0.05 &  3.34 $\pm$ 0.03\\   
TWA 11B        &   M2.5\tablenotemark{b} &  5 &  0.503 $\pm$   0.022 &   3653 $\pm$    14 & 4.26 $\pm$  0.05 &  3.22 $\pm$ 0.03\\   
TWA 12         &   M2\tablenotemark{b}   &  5 &  0.521 $\pm$   0.020 &   3647 $\pm$    13 & 4.33 $\pm$  0.05 &  3.17 $\pm$ 0.03\\   
TWA 13A        &   M1e\tablenotemark{b}  &  5 &  0.533 $\pm$   0.021 &   3845 $\pm$    16 & 4.46 $\pm$  0.05 &  3.23 $\pm$ 0.03\\   
TWA 13B        &   M2e\tablenotemark{b}  &  5 &  0.577 $\pm$   0.021 &   3817 $\pm$    16 & 4.47 $\pm$  0.05 &  3.36 $\pm$ 0.03\\   
TWA 14         &   M0\tablenotemark{b}   &  5 &  0.589 $\pm$   0.021 &   3701 $\pm$    13 & 4.45 $\pm$  0.05 &  3.63 $\pm$ 0.06\\   
TWA 15A        &   M1.5\tablenotemark{b} &  5 &  0.494 $\pm$   0.025 &   3482 $\pm$    12 & 4.32 $\pm$  0.05 &  3.07 $\pm$ 0.05\\   
TWA 15B        &   M2\tablenotemark{b}   &  5 &  0.484 $\pm$   0.030 &   3483 $\pm$    12 & 4.30 $\pm$  0.06 &  3.16 $\pm$ 0.05\\   
TWA 16         &   M1.5\tablenotemark{b} &  5 &  0.354 $\pm$   0.024 &   3610 $\pm$    20 & 4.27 $\pm$  0.05 &  2.55 $\pm$ 0.04\\   
TWA 17         &   K5\tablenotemark{b}   &  5 &  0.499 $\pm$   0.021 &   3884 $\pm$    25 & 4.02 $\pm$  0.06 &  3.31 $\pm$ 0.10\\   
TWA 18         &   M0.5\tablenotemark{b} &  5 &  0.464 $\pm$   0.021 &   3780 $\pm$    13 & 4.50 $\pm$  0.06 &  3.14 $\pm$ 0.04\\   
TWA 19A        &   G5\tablenotemark{a}   &  5 &  0.191 $\pm$   0.020 &   5986 $\pm$    14 & 3.83 $\pm$  0.07 &  3.92 $\pm$ 0.03\\   
TWA 19B        &   K7\tablenotemark{b}   &  5 &  0.452 $\pm$   0.022 &   3896 $\pm$    15 & 4.24 $\pm$  0.05 &  2.89 $\pm$ 0.07\\   
TWA 21         &   K3/4\tablenotemark{a} &  5 &  0.369 $\pm$   0.020 &   4689 $\pm$    14 & 4.44 $\pm$  0.05 &  3.26 $\pm$ 0.03\\   
TWA 22         &   M5\tablenotemark{a}   &  5 &  0.616 $\pm$   0.021 &   2990 $\pm$    13 & 4.20 $\pm$  0.05 &  3.57 $\pm$ 0.06\\   
TWA 23         &   M1\tablenotemark{a}   &  5 &  0.525 $\pm$   0.020 &   3466 $\pm$    12 & 4.12 $\pm$  0.05 &  3.51 $\pm$ 0.04\\   
TWA 24N        &                &  5 &  0.438 $\pm$   0.032 &   3669 $\pm$    15 & 4.28 $\pm$  0.06 &  2.83 $\pm$ 0.08\\   
TWA 24S        &   K3\tablenotemark{a}   &  5 &  0.380 $\pm$   0.020 &   4920 $\pm$    16 & 4.32 $\pm$  0.05 &  3.64 $\pm$ 0.02\\   
TWA 25         &   M0\tablenotemark{a}   &  1 &  0.570 $\pm$   0.020 &   3920 $\pm$    10 & 4.45 $\pm$  0.05 &  3.25 $\pm$ 0.02\\   

\enddata
\tablecomments{all uncertainties are internal, derived from the scatter of fitted values from individual spectra from the mean. The external uncertainties are much larger (see \S\ref{s:LiRes}).}
\tablenotetext{a}{\citet{zuc04}}
\tablenotetext{b}{\citet{del06}}
\label{tab:TWA}
\end{deluxetable}%

\begin{deluxetable}{llcllll}
\tablecaption{Results for stars in $\eta$~Chamaeleontis}
\tablecolumns{7}
\tablewidth{0pc}
\tabletypesize{\small}
\tablehead{
	\colhead{Object ID} 			&
	\colhead{Sp.T.} 			&
	\colhead{\# Obs.}			&
	\colhead{EW$_{6708}$ (\AA)} &
	\colhead{$T_{\mathrm{eff}}$ (K)} 	&
	\colhead{$\log g$ }			&
	\colhead{$\log N_{\mathrm{Li}}$}		
}
\startdata
$\eta$\,Cha 1    &  K4\tablenotemark{a}    &  5 &  0.511 $\pm$    0.020  &  4107 $\pm$    15 & 3.94 $\pm$  0.05 &  3.31 $\pm$ 0.05 \\
$\eta$\,Cha 3    &  M3.25\tablenotemark{b} &  5 &  0.542 $\pm$    0.021  &  3508 $\pm$    12 & 4.18 $\pm$  0.05 &  3.55 $\pm$ 0.04 \\   
$\eta$\,Cha 4    &  K7\tablenotemark{a}    &  5 &  0.587 $\pm$    0.021  &  3822 $\pm$    11 & 4.43 $\pm$  0.05 &  3.32 $\pm$ 0.03 \\   
$\eta$\,Cha 5    &  M4\tablenotemark{b}    &  5 &  0.606 $\pm$    0.021  &  3314 $\pm$    11 & 4.04 $\pm$  0.05 &  3.41 $\pm$ 0.09 \\   
$\eta$\,Cha 6    &  M2\tablenotemark{a}    &  4 &  0.489 $\pm$    0.022  &  3492 $\pm$    11 & 4.22 $\pm$  0.06 &  3.31 $\pm$ 0.03 \\   
$\eta$\,Cha 7    &  K6\tablenotemark{b}    &  5 &  0.415 $\pm$    0.022  &  4002 $\pm$    25 & 3.81 $\pm$  0.05 &  2.77 $\pm$ 0.08 \\   
$\eta$\,Cha 9    &  M4.5\tablenotemark{a}  &  5 &  0.566 $\pm$    0.027  &  3127 $\pm$    11 & 3.84 $\pm$  0.05 &  3.36 $\pm$ 0.11 \\ 
$\eta$\,Cha 10   &  K7\tablenotemark{a}    &  5 &  0.534 $\pm$    0.021  &  3992 $\pm$    11 & 4.67 $\pm$  0.06 &  3.08 $\pm$ 0.03 \\   
$\eta$\,Cha 11   &  K4\tablenotemark{a}    &  5 &  0.483 $\pm$    0.021  &  4182 $\pm$    19 & 3.92 $\pm$  0.05 &  3.28 $\pm$ 0.03 \\   
$\eta$\,Cha 12   &  M2\tablenotemark{a}    &  5 &  0.609 $\pm$    0.023  &  3440 $\pm$    11 & 4.31 $\pm$  0.06 &  3.20 $\pm$ 0.08 \\   
$\eta$\,Cha 13   &  M2\tablenotemark{a}    &  5 &  0.426 $\pm$    0.023  &  3139 $\pm$    11 & 3.68 $\pm$  0.05 &  2.89 $\pm$ 0.07 \\

\enddata
\tablecomments{all uncertainties are internal, derived from the scatter of fitted values from individual spectra from the mean. The external uncertainties are much larger (see \S\ref{s:LiRes}).}
\tablenotetext{a}{\citet{zuc04}}
\tablenotetext{b}{\citet{luh04}}
\label{tab:etaCha}
\end{deluxetable}%

\begin{deluxetable}{llcllll}
\tablecaption{Results for stars in the $\beta$~Pictoris moving group (BPMG)}
\tablecolumns{7}
\tablewidth{0pc}
\tabletypesize{\small}
\tablehead{
	\colhead{Object ID} 			&
	\colhead{Sp.T.} 			&
	\colhead{\# Obs.}			&
	\colhead{EW$_{6708}$ (\AA)} &
	\colhead{$T_{\mathrm{eff}}$ (K)} 	&
	\colhead{$\log g$ }			&
	\colhead{$\log N_{\mathrm{Li}}$}		
}
\startdata
AO Men          &  K6/7         &   5 &  0.409 $\pm$   0.020  &  4317 $\pm$    12 & 4.13 $\pm$  0.05 &  3.01 $\pm$ 0.02 \\
AU Mic          &  M1e          &   5 &  0.067 $\pm$   0.020  &  3841 $\pm$    11 & 4.66 $\pm$  0.05 &  0.80 $\pm$ 0.03 \\   
BD -17 6128A    &  K7e/M0       &   5 &  0.427 $\pm$   0.020  &  4140 $\pm$    13 & 4.37 $\pm$  0.05 &  2.78 $\pm$ 0.02 \\   
BD -17 6128B    &               &   4 &  0.051 $\pm$   0.035  &  3350 $\pm$    19 & 4.28 $\pm$  0.05 &  0.54 $\pm$ 0.10 \\   
CD -64 1208     &  K7           &   5 &  0.464 $\pm$   0.021  &  3740 $\pm$    19 & 3.96 $\pm$  0.05 &  1.27 $\pm$ 0.04 \\   
GJ 3305         &  M0.5         &   5 &  0.077 $\pm$   0.020  &  3840 $\pm$    11 & 4.72 $\pm$  0.05 &  1.09 $\pm$ 0.03 \\   
GJ 799N         &  M4.5e        &   5 &  0.020 $\pm$   0.020  &  3237 $\pm$    10 & 4.47 $\pm$  0.05 &  0.25 $\pm$ 0.02 \\   
GJ 799S         &  M4.5e        &   5 &  0.020 $\pm$   0.021  &  3244 $\pm$    11 & 4.46 $\pm$  0.05 &  0.25 $\pm$ 0.02 \\   
HD 164249B      &               &   4 &  0.075 $\pm$   0.029  &  3505 $\pm$    18 & 4.32 $\pm$  0.10 &  0.25 $\pm$ 0.10 \\   
HD 181327       &  F5.5         &   5 &  0.122 $\pm$   0.020  &  6392 $\pm$    11 & 4.17 $\pm$  0.08 &  3.70 $\pm$ 0.02 \\   
HD 35850        &  F7           &   5 &  0.146 $\pm$   0.020  &  6275 $\pm$    18 & 4.15 $\pm$  0.08 &  3.68 $\pm$ 0.02 \\   
HIP 10679       &  G2V          &   1 &  0.172 $\pm$   0.020  &  6080 $\pm$    10 & 4.25 $\pm$  0.05 &  3.70 $\pm$ 0.02 \\   
HIP 10680       &  F5V          &   1 &  0.132 $\pm$   0.020  &  6250 $\pm$    10 & 4.00 $\pm$  0.05 &  3.65 $\pm$ 0.02 \\   
HIP 112312A     &  M4e          &   3 &  0.020 $\pm$   0.020  &  3340 $\pm$    12 & 4.37 $\pm$  0.07 &  0.25 $\pm$ 0.02 \\   
HIP 112312B     &  M4.5         &   3 &  0.315 $\pm$   0.022  &  3170 $\pm$    14 & 4.17 $\pm$  0.07 &  2.17 $\pm$ 0.19 \\   
HIP 11437A      &  K8           &   1 &  0.248 $\pm$   0.020  &  4310 $\pm$    10 & 4.60 $\pm$  0.05 &  2.30 $\pm$ 0.02 \\   
HIP 11437B      &  M0           &   1 &  0.130 $\pm$   0.020  &  3620 $\pm$    10 & 4.55 $\pm$  0.05 &  1.30 $\pm$ 0.02 \\   
HIP 12545       &  M0           &   2 &  0.433 $\pm$   0.023  &  4205 $\pm$    11 & 4.22 $\pm$  0.06 &  2.92 $\pm$ 0.03 \\   
HIP 23309       &  M5           &   5 &  0.336 $\pm$   0.022  &  4041 $\pm$    10 & 4.72 $\pm$  0.05 &  2.46 $\pm$ 0.02 \\   
HIP 23418N      &  M3V          &   2 &  0.020 $\pm$   0.021  &  3335 $\pm$    11 & 4.43 $\pm$  0.06 &  0.25 $\pm$ 0.02 \\   
HIP 23418S      &               &   2 &  0.020 $\pm$   0.022  &  3310 $\pm$    14 & 4.35 $\pm$  0.07 &  0.25 $\pm$ 0.02 \\   
PZ Tel          &  K0Vp         &   5 &  0.271 $\pm$   0.020  &  5537 $\pm$    15 & 3.98 $\pm$  0.11 &  3.65 $\pm$ 0.02 \\   
V343 Nor        &  K0V          &   5 &  0.286 $\pm$   0.020  &  5520 $\pm$    13 & 4.22 $\pm$  0.07 &  3.75 $\pm$ 0.02 \\   

\enddata
\tablecomments{all uncertainties are internal, derived from the scatter of fitted values from individual spectra from the mean. The external uncertainties are much larger (see \S\ref{s:LiRes}). All spectral types from \citet{zuc04}}
\label{tab:BPMG}
\end{deluxetable}%

\begin{deluxetable}{llcllll}
\tablecaption{Results for stars in Tucanae--Horologium (TUCHOR)}
\tablecolumns{7}
\tablewidth{0pc}
\tabletypesize{\small}
\tablehead{
	\colhead{Object ID} 			&
	\colhead{Sp.T.} 			&
	\colhead{\# Obs.}			&
	\colhead{EW$_{6708}$ (\AA)} &
	\colhead{$T_{\mathrm{eff}}$ (K)} 	&
	\colhead{$\log g$ }			&
	\colhead{$\log N_{\mathrm{Li}}$}		
}
\startdata
CD -53\,544       &  K6Ve         &   5 &  0.275 $\pm$   0.021  &  3796  $\pm$   13 & 4.53 $\pm$  0.05 &  1.86 $\pm$ 0.04 \\
CD -60\,416       &  K3/4         &   5 &  0.237 $\pm$   0.021  &  4310  $\pm$   33 & 4.34 $\pm$  0.06 &  2.33 $\pm$ 0.03 \\   
CPD -64\,120      &  K1Ve         &   5 &  0.284 $\pm$   0.023  &  5212  $\pm$   48 & 4.61 $\pm$  0.05 &  3.73 $\pm$ 0.03 \\   
GSC 8056-0482   &  M3Ve         &   5 &  0.353 $\pm$   0.021  &  3541  $\pm$   13 & 4.54 $\pm$  0.05 &  2.38 $\pm$ 0.06 \\   
GSC 8491-1194   &  M3Ve         &   5 &  0.033 $\pm$   0.021  &  3578  $\pm$   12 & 4.64 $\pm$  0.06 &  0.25 $\pm$ 0.02 \\   
GSC 8497-0995   &  K6Ve         &   4 &  0.127 $\pm$   0.022  &  4238  $\pm$   18 & 4.64 $\pm$  0.07 &  1.52 $\pm$ 0.05 \\   
HD 13183        &  G5V          &   5 &  0.205 $\pm$   0.020  &  5854  $\pm$   22 & 4.20 $\pm$  0.05 &  3.73 $\pm$ 0.02 \\   
HD 13246        &  F8V          &   5 &  0.133 $\pm$   0.020  &  6292  $\pm$   13 & 4.17 $\pm$  0.06 &  3.68 $\pm$ 0.03 \\   
HD 8558         &  G6V          &   5 &  0.192 $\pm$   0.020  &  5825  $\pm$   19 & 4.27 $\pm$  0.05 &  3.90 $\pm$ 0.02 \\   
HD 9054         &  K1V          &   5 &  0.178 $\pm$   0.020  &  5165  $\pm$   35 & 4.65 $\pm$  0.05 &  3.01 $\pm$ 0.02 \\   
HIP 105388      &  G5V          &   5 &  0.224 $\pm$   0.020  &  5748  $\pm$   11 & 4.18 $\pm$  0.08 &  3.71 $\pm$ 0.02 \\   
HIP 105404      &  K0V          &   5 &  0.171 $\pm$   0.022  &  5543  $\pm$   54 & 4.28 $\pm$  0.10 &  3.56 $\pm$ 0.05 \\   
HIP 107345      &  M1           &   5 &  0.055 $\pm$   0.021  &  3975  $\pm$   13 & 4.82 $\pm$  0.05 &  0.25 $\pm$ 0.09 \\   
HIP 108422      &  G8V          &   5 &  0.261 $\pm$   0.020  &  5541  $\pm$   27 & 3.99 $\pm$  0.07 &  N/A\tablenotemark{a}\\   
HIP 1113        &  G6V          &   3 &  0.202 $\pm$   0.020  &  5757  $\pm$   13 & 4.35 $\pm$  0.07 &  3.80 $\pm$ 0.04 \\   
HIP 116748N     &              &   3 &  0.218 $\pm$   0.020  &  4620  $\pm$   10 & 4.47 $\pm$  0.05 &  2.62 $\pm$ 0.03 \\   
HIP 116749S     &  G51V         &   3 &  0.212 $\pm$   0.021  &  5813  $\pm$   44 & 4.30 $\pm$  0.06 &  4.00 $\pm$ 0.02 \\   
HIP 1481        &  F8           &   3 &  0.128 $\pm$   0.020  &  6323  $\pm$   13 & 4.27 $\pm$  0.06 &  3.68 $\pm$ 0.03 \\   
HIP 16853       &  G2V          &   5 &  0.149 $\pm$   0.020  &  6217  $\pm$   14 & 4.29 $\pm$  0.07 &  4.00 $\pm$ 0.02 \\   
HIP 1910        &  M1           &   3 &  0.181 $\pm$   0.020  &  3890  $\pm$   20 & 4.73 $\pm$  0.06 &  1.97 $\pm$ 0.06 \\   
HIP 1993        &  M1           &   3 &  0.038 $\pm$   0.022  &  4017  $\pm$   16 & 4.83 $\pm$  0.08 &  0.25 $\pm$ 0.02 \\   
HIP 21632       &  G3V          &   5 &  0.188 $\pm$   0.020  &  6003  $\pm$   18 & 4.12 $\pm$  0.06 &  3.82 $\pm$ 0.02 \\   
HIP 22295       &  F7V          &   5 &  0.130 $\pm$   0.020  &  6342  $\pm$   17 & 4.13 $\pm$  0.09 &  3.65 $\pm$ 0.02 \\   
HIP 2729        &  K5V          &   3 &  0.338 $\pm$   0.020  &  3827  $\pm$   11 & 4.02 $\pm$  0.06 &  0.68 $\pm$ 0.03 \\   
HIP 30030       &  G0           &   5 &  0.163 $\pm$   0.021  &  6210  $\pm$   21 & 4.19 $\pm$  0.08 &  3.77 $\pm$ 0.04 \\   
HIP 30034       &  K2V          &   5 &  0.287 $\pm$   0.020  &  5268  $\pm$   23 & 4.69 $\pm$  0.06 &  3.57 $\pm$ 0.02 \\   
HIP 32235       &  G6V          &   5 &  0.233 $\pm$   0.020  &  5774  $\pm$   13 & 4.34 $\pm$  0.07 &  3.99 $\pm$ 0.02 \\   
HIP 33737       &  K3V          &   5 &  0.279 $\pm$   0.020  &  4859  $\pm$   18 & 4.64 $\pm$  0.05 &  3.16 $\pm$ 0.03 \\   
HIP 3556        &  M3           &   3 &  0.055 $\pm$   0.022  &  3677  $\pm$   17 & 4.72 $\pm$  0.05 &  0.27 $\pm$ 0.09 \\   
HIP 490         &  G0V          &   3 &  0.153 $\pm$   0.020  &  6173  $\pm$   17 & 4.28 $\pm$  0.05 &  4.00 $\pm$ 0.02 \\   
HIP 9141        &  G3/5V        &   5 &  0.187 $\pm$   0.020  &  5992  $\pm$   13 & 4.23 $\pm$  0.05 &  3.80 $\pm$ 0.02 \\   
TUCh 7600-0516  &  K1           &   5 &  0.249 $\pm$   0.020  &  5163  $\pm$   31 & 4.63 $\pm$  0.06 &  3.56 $\pm$ 0.02 \\   
TYC 5882-1169   &  K3/4         &   5 &  0.241 $\pm$   0.022  &  4939  $\pm$   34 & 4.65 $\pm$  0.09 &  3.05 $\pm$ 0.11 \\   
TYC 7065-0879N     &  K4           &   3 &  0.259 $\pm$   0.020  &  5513  $\pm$   28 & 4.58 $\pm$  0.07 &  3.63 $\pm$ 0.03 \\   
TYC 7065-0879S     &  K4           &   3 &  0.267 $\pm$   0.021  &  5453  $\pm$   31 & 4.70 $\pm$  0.14 &  3.63 $\pm$ 0.03 \\   

\enddata
\tablenotetext{a}{Lithium abundance could not be measured due to high rotational broadening in HIP\,108422 (v$\sin$\,i = 139.80 km/s)}
\tablecomments{all uncertainties are internal, derived from the scatter of fitted values from individual spectra from the mean. The external uncertainties are much larger (see \S\ref{s:LiRes}). All spectral types from \citet{zuc04}}
\label{tab:TH}
\end{deluxetable}%

\begin{deluxetable}{llcllll}
\tablecaption{Results for stars in AB Doradus (ABD)}
\tablecolumns{7}
\tablewidth{0pc}
\tabletypesize{\small}
\tablehead{
	\colhead{Object ID} 			&
	\colhead{Sp.T.} 			&
	\colhead{\# Obs.}			&
	\colhead{EW$_{6708}$ (\AA)} &
	\colhead{$T_{\mathrm{eff}}$ (K)} 	&
	\colhead{$\log g$ }			&
	\colhead{$\log N_{\mathrm{Li}}$}		
}
\startdata
AB Dor          &  K1            &  3 &  0.261 $\pm$   0.021 &   5210 $\pm$    34 & 4.63 $\pm$  0.06 &  3.50 $\pm$ 0.04 \\
GSC 08894-00426 &  M2            &  3 &  0.037 $\pm$   0.022 &   3343 $\pm$    11 & 4.83 $\pm$  0.05 &  0.25 $\pm$ 0.02 \\   
HD 13482A       &  K1            &  2 &  0.124 $\pm$   0.020 &   5725 $\pm$   105 & 4.47 $\pm$  0.18 &  3.17 $\pm$ 0.03 \\   
HD 13482B       &                &  1 &  0.076 $\pm$   0.020 &   5090 $\pm$    10 & 4.95 $\pm$  0.05 &  1.75 $\pm$ 0.02 \\   
HD 17332B       &                &  1 &  0.168 $\pm$   0.020 &   5780 $\pm$    10 & 4.35 $\pm$  0.05 &  3.55 $\pm$ 0.02 \\   
HD 217343       &  G3V           &  3 &  0.172 $\pm$   0.020 &   5957 $\pm$    28 & 4.25 $\pm$  0.05 &  3.68 $\pm$ 0.04 \\   
HD 217379N      &                &  2 &  0.020 $\pm$   0.020 &   4050 $\pm$    14 & 5.07 $\pm$  0.06 &  0.30 $\pm$ 0.05 \\   
HD 217379S      &                &  2 &  0.030 $\pm$   0.021 &   4165 $\pm$    11 & 5.05 $\pm$  0.07 &  0.35 $\pm$ 0.05 \\   
HD 218860       &  G5            &  2 &  0.222 $\pm$   0.020 &   5670 $\pm$    10 & 4.32 $\pm$  0.06 &  3.55 $\pm$ 0.02 \\   
HD 224228       &  K3V           &  2 &  0.076 $\pm$   0.020 &   4905 $\pm$    11 & 4.82 $\pm$  0.06 &  1.35 $\pm$ 0.02 \\   
HD 35650        &  K7            &  4 &  0.020 $\pm$   0.020 &   4252 $\pm$    10 & 5.06 $\pm$  0.05 &  0.25 $\pm$ 0.02 \\   
HD 45270        &  G1V           &  3 &  0.142 $\pm$   0.020 &   6187 $\pm$    24 & 4.50 $\pm$  0.10 &  3.90 $\pm$ 0.04 \\   
HD 65569        &  K1V           &  3 &  0.155 $\pm$   0.020 &   5223 $\pm$    31 & 4.87 $\pm$  0.05 &  2.77 $\pm$ 0.03 \\   
HIP 14807       &  K6            &  1 &  0.034 $\pm$   0.020 &   4500 $\pm$    10 & 4.60 $\pm$  0.05 &  0.55 $\pm$ 0.02 \\   
HIP 14809       &  G5            &  1 &  0.150 $\pm$   0.020 &   6050 $\pm$    10 & 4.25 $\pm$  0.05 &  3.60 $\pm$ 0.02 \\   
HIP 17695       &  M3            &  2 &  0.074 $\pm$   0.029 &   3545 $\pm$    18 & 4.72 $\pm$  0.06 &  0.25 $\pm$ 0.02 \\   
HIP 26369       &  K7            &  3 &  0.044 $\pm$   0.020 &   4010 $\pm$    15 & 4.55 $\pm$  0.06 &  0.25 $\pm$ 0.02 \\   
HIP 31878       &  K7            &  3 &  0.042 $\pm$   0.024 &   4240 $\pm$    14 & 5.12 $\pm$  0.05 &  0.53 $\pm$ 0.03 \\   
HIP 6276        &  G8            &  3 &  0.153 $\pm$   0.020 &   5643 $\pm$    13 & 4.43 $\pm$  0.05 &  3.22 $\pm$ 0.04 \\   
HR 2468         &  G1.5          &  2 &  0.138 $\pm$   0.020 &   6060 $\pm$    14 & 4.60 $\pm$  0.11 &  3.58 $\pm$ 0.03 \\   
UY Pic          &  K0V           &  3 &  0.267 $\pm$   0.020 &   5600 $\pm$    15 & 5.05 $\pm$  0.06 &  3.60 $\pm$ 0.02 \\   
V372 Pub        &  M3            &  5 &  0.020 $\pm$   0.021 &   3688 $\pm$    19 & 4.88 $\pm$  0.05 &  0.25 $\pm$ 0.02 \\   

\enddata
\tablecomments{all uncertainties are internal, derived from the scatter of fitted values from individual spectra from the mean. The external uncertainties are much larger (see \S\ref{s:LiRes}). All spectral types from \citet{zuc04}}
\label{tab:ABD}
\end{deluxetable}%

\begin{deluxetable}{llcllll}
\tablecaption{Results for the field stars (radial velocity standards)}
\tablecolumns{7}
\tablewidth{0pc}
\tabletypesize{\small}
\tablehead{
	\colhead{Object ID} 			&
	\colhead{Sp.T.} 			&
	\colhead{\# Obs.}			&
	\colhead{$T_{\mathrm{eff}}$ (K)} 	&
	\colhead{$\log g$ }			&
	\colhead{$\log N_{\mathrm{Li}}$}		
}
\startdata
GJ 729     &  M3.5          &  1 &  3340 $\pm$    10 & 5.00 $\pm$  0.05 &  0.25 $\pm$ 0.02 \\
GJ 156     &  K7            &  4 &  4110 $\pm$    11 & 5.19 $\pm$  0.05 &  0.25 $\pm$ 0.02 \\   
Gl 205     &  M1.5          &  4 &  3875 $\pm$    10 & 4.89 $\pm$  0.06 &  0.25 $\pm$ 0.02 \\   
Gl 349     &  K3            &  5 &  4712 $\pm$    24 & 4.79 $\pm$  0.07 &  0.35 $\pm$ 0.02 \\   
Gl 382     &  M1.5          &  2 &  3745 $\pm$    11 & 4.68 $\pm$  0.06 &  0.25 $\pm$ 0.03 \\   
Gl 876     &  M4            &  1 &  3290 $\pm$    10 & 4.60 $\pm$  0.05 &  0.25 $\pm$ 0.02 \\   
Gl 880     &  M1.5          &  2 &  3795 $\pm$    11 & 4.97 $\pm$  0.06 &  0.25 $\pm$ 0.02 \\   
HD 103932  &  K5            &  5 &  4360 $\pm$    10 & 4.77 $\pm$  0.05 &  0.25 $\pm$ 0.02 \\   
HD 111631  &  K7            &  5 &  4073 $\pm$    10 & 5.19 $\pm$  0.05 &  0.25 $\pm$ 0.02 \\   
HD 153458  &  G0            &  2 &  5935 $\pm$    11 & 4.12 $\pm$  0.06 &  2.25 $\pm$ 0.02 \\ 
HD 172051  &  G5V           &  3 &  6070 $\pm$    12 & 4.45 $\pm$  0.06 &  0.97 $\pm$ 0.03 \\  
HD 120467  &  K4            &  5 &  4231 $\pm$    11 & 4.97 $\pm$  0.05 &  0.25 $\pm$ 0.02 \\   
HD 83443   &  K0            &  3 &  5280 $\pm$    12 & 3.90 $\pm$  0.06 &  0.45 $\pm$ 0.02 \\   
HD 87359   &  G5            &  3 &  5903 $\pm$    35 & 4.15 $\pm$  0.05 &  0.92 $\pm$ 0.04 \\   
HD 88218   &  F8            &  1 &  6240 $\pm$    10 & 4.10 $\pm$  0.05 &  2.80 $\pm$ 0.02 \\   
HD 92945   &  K1            &  1 &  5310 $\pm$    10 & 4.70 $\pm$  0.05 &  2.95 $\pm$ 0.02 \\   
HD 96700   &  G2            &  1 &  6160 $\pm$    10 & 4.20 $\pm$  0.05 &  1.55 $\pm$ 0.02 \\   
LHS 1763   &  K5            &  4 &  4348 $\pm$    10 & 4.78 $\pm$  0.05 &  0.49 $\pm$ 0.02 \\   
NSV 2863   &  M1.5          &  4 &  3882 $\pm$    11 & 4.94 $\pm$  0.06 &  0.25 $\pm$ 0.02 \\   
NSV 6431   &  M2            &  5 &  3724 $\pm$    10 & 4.64 $\pm$  0.05 &  0.25 $\pm$ 0.02 \\   

\enddata
\tablecomments{all uncertainties are internal, derived from the scatter of fitted values from individual spectra from the mean. The external uncertainties are much larger (see \S\ref{s:LiRes}).}
\label{tab:RVstd}
\end{deluxetable}%

\end{document}